\documentclass[11pt]{article}
\usepackage{latexsym} \usepackage{epsf}
\usepackage{a4}
\usepackage{amsfonts}
\renewcommand{\theequation}{\thesection.\arabic{equation}}
\newcounter{subequation}[equation]
\makeatletter

\expandafter\let\expandafter\reset@font\csname reset@font\endcsname

\def\subeqnarray{\arraycolsep1pt
    \def\@eqnnum\stepcounter##1{\stepcounter{subequation}%
	{\reset@font\rm(\theequation\alph{subequation})}}
\jot5mm     \eqnarray}

\makeatother

\def\rmd{\hbox{\rm d}}
\def\u{\hbox{\rm u}}

\def\be{\begin{equation}}
\def\ee{\end{equation}}
\def\bea{\begin{eqnarray}}
\def\eea{\end{eqnarray}}
\def\dd{\partial}
\def\half{\frac{1}{2}}

\def\one#1{#1^{\raise5pt\hbox{$\scriptstyle\!\!\!\!1$}}\,{}}
\def\two#1{#1^{\raise5pt\hbox{$\scriptstyle\!\!\!\!2$}}\,{}}

\makeatletter
\def\binrel@#1{\begingroup
  \setboxz@h{\thinmuskip0mu
    \medmuskip\m@ne mu\thickmuskip\@ne mu
    \setbox\tw@\hbox{$#1\m@th$}\kern-\wd\tw@
    ${}#1{}\m@th$}%
  \edef\@tempa{\endgroup\let\noexpand\binrel@@
    \ifdim\wdz@<\z@ \mathbin
    \else\ifdim\wdz@>\z@ \mathrel
    \else \relax\fi\fi}%
  \@tempa
}
\let\binrel@@\relax
\def\overset#1#2{\binrel@{#2}%
  \binrel@@{\mathop{\kern\z@#2}\limits^{#1}}}
\def\underset#1#2{\binrel@{#2}%
  \binrel@@{\mathop{\kern\z@#2}\limits_{#1}}}
\makeatother
\newfont{\bbd}{msbm10 scaled\magstep1}
\def\C{\hbox{\bbd C}}
\def\B{\hbox{\bbd B}}
\def\D{\hbox{\bbd D}}

\def\R{\hbox{\bbd R}}
\def\V{\hbox{\bbd V}}
\def\W{\hbox{\bbd W}}
\def\S{\hbox{\bbd S}}

\def\P{\hbox{\bbd P}}

\def\Z{{\mathcal Z}}
\def\DD{{\mathcal D}}

\def\RR{{\mathcal R}}

\newtheorem{prop}{Proposition}

\begin{document}
\hfill NTZ 1/2001

{\begin{center}
{\LARGE {Universal $\R$-matrix as integral operator.} } \\ [8mm]
{\large  S.Derkachov$^{a,c}$\footnote{e-mail: 
Sergey.Derkachov@itp.uni-leipzig.de}, 
D. Karakhanyan$^b$\footnote{e-mail: karakhan@lx2.yerphi.am} \& 
R. Kirschner$^c$\footnote{e-mail:Roland.Kirschner@itp.uni-leipzig.de} \\ [3mm] } 
\end{center} 
} 

\begin{itemize}
\item[$^a$] 
Department of Mathematics, St Petersburg Technology Institute, \\
Sankt Petersburg, Russia   
\item[$^b$] 
Yerevan Physics Institute , \\
Br.Alikhanian st.2, 375036, Yerevan , Armenia. 
\item[$^c$] Naturwissenschaftlich-Theoretisches Zentrum und 
Institut f\"{u}r Theoretische Physik, Universit\"{a}t Leipzig, \\
Augustusplatz 10, D-04109 Leipzig, Germany
\end{itemize}

\vspace{3cm} 
\noindent 
{\bf Abstract.}   

We derive the integral operator form for 
the general rational solution of the Yang-Baxter 
equation with $s\ell(2|1)$ symmetry. 
Considering the defining relations for 
the kernel of the R-operator as a 
system of second order differential equations 
we observe remarkable reduction to a system 
of simple first order equations.
The obtained kernel of R-operator has a very simple structure.
To illustrate all this in the simplest situation 
we treat also the $s\ell(2)$ case.

\newpage


{\small \tableofcontents}
\renewcommand{\refname}{References.}
\renewcommand{\thefootnote}{\arabic{footnote}}
\setcounter{footnote}{0} 
\setcounter{equation}{0}

\renewcommand{\theequation}{\thesubsection.\arabic{equation}}
\setcounter{equation}{0}

\section{Introduction}
\setcounter{equation}{0}

Solutions of the Yang-Baxter equation, in particular ones with $s\ell(2)$ or
$s\ell(2|1) $ symmetry, are used in constructing a number of 
statistical models~\cite{BIK,Maas,RM,EK,FK,PF,A,K}
and are applied in analyzing the 
high-energy Regge~\cite{L,FadK,Kor,KK}
and Bjorken asymptotics~\cite{BDM,DKM,B}
of four -dimensional gauge theories.

In the previous paper~\cite{DKK} we have studied the $s\ell(2|1)$ symmetric
rational solutions of the Yang-Baxter equation. We have obtained the
general R-operator acting on the tensor product of two
arbitrary representation spaces of lowest weight in terms of the matrix
elements in a basis of lowest weight vectors appearing in the tensor
product.

The lowest weight representation spaces have been chosen in terms of
polynomials in one even variable $z$ and two 
odd variables $\theta, \overline \theta$. 
Then the tensor product representation consists of polynomial
two-point functions, i.e. functions of two sets $ (z_i, \theta_i,
\overline \theta_i) $. This formulation is the appropriate one for
treating arbitrary infinite dimensional lowest weight representations.
It is the natural formulation from the viewpoint of applications 
to the Bjorken limit of four-dimensional field theories like QCD.
In the Bjorken asymtotics the interaction reduces in
dimension effectively  to a line on the light cone, being the range of
the even variable $z$. The Grassmann variables $\theta, \overline
\theta$ appear in the Bjorken limit of supersymmetric (${\cal N} = 1 $)
Yang-Mills theories allowing to unify the gluon and quark fields in
terms of superfields. The effective interaction can be formulated in
terms of integral operators, the kernels of which are 
(super)conformal four-point functions.

This physical context provides a motivation to look for a corresponding
formulation of the R-operators in terms of integral kernels. Instead of
constructing these kernels from the matrix elements derived earlier we
present here a direct calculation. We consider the defining relations
for the general R-operator now as differential equations on the kernel.
The essential observation is that this set of differential equation
reduces to  simple first order differential equations, one for each
variable involved. The resulting kernel is expressed in powers of
combinations of the variables having simple $s\ell(2|1) $ transformation
properties.

The method proposed here seems to be general enough to
become useful in solving the Yang-Baxter equation with other symmetries.
In particular the observed reduction of the defining relation to simple
first order differential equations is expected to work as a powerful
tool in other cases. This reduction relies on the structure of the Lax
operator, i.e. the R-operator acting in the tensor product of the
fundamental and an arbitrary representation. Its matrix representation
decomposes into matrices times derivatives, where these matrices are
tensor products of particular vectors.
In the case of $s\ell(2)$ this structure appeared earlier in calculations
by L.N.Lipatov~\cite{L} and has been used in \cite{FadK,Kor,KK}.
The kernel of R-operator is finally obtained as the similarity 
transformation transforming these first order differential operators 
into the simple derivatives with respect to 
variables~$z_i,\theta_i,\bar\theta_i$.    

Aiming to give a clear presentation of the method we 
discuss the simplest case of $s\ell(2)$ first.
Note that the main idea to represent the universal 
R-matrix as an integral operator originates from the paper~\cite{Skl}.
Moreover, the final answer for the kernel of R-operator 
has been obtained by a straightforward solution of the 
system of differential equations of the second order 
in the paper~\cite{DKS}. See this paper for the applications of the 
integral representation of R-matrix. 
 
The presentation is organized as follows.
In Section 2 we collect the standard facts 
about the algebra~$s\ell(2)$ and its 
representations.
We represent the lowest weight modules by polynomials in one 
variable~($z$) and the $s\ell(2)$-generators as first order 
differential operators.
We derive the defining relation for 
the general R-matrix, i.e. the solution of the Yang-Baxter equation 
acting on tensor products of two arbitrary representations, 
the elements of which are polynomial functions 
of~$z_1$ and~$z_2$. We solve this defining relation in 
the space of lowest weights.
The result is well known~\cite{F,KRS,KS,H} but we represent 
all calculations in the form which is 
appropriate for the supersymmetric generalisation.
In Section 3 we give the simple method of derivation 
of the integral operator representation for the R-matrix.
The main observation is that the defining system of 
differential equations of the second order for 
the kernel of R-matrix is equivalent to the 
system of the first-order differential equations.   

These sections have introductory character and 
starting from the Section 4 we follow the same strategy 
for the supersymmetric algebra~$s\ell(2|1)$.
We represent the~$s\ell(2|1)$ lowest weight modules by polynomials in one 
even~($z$) and two odd variables~($\theta,\bar\theta$) and 
the $s\ell(2|1)$-generators as first order 
differential operators and derive the defining relation for 
the general R-matrix, i.e. the solution of the Yang-Baxter equation 
acting on tensor products of two arbitrary representations, 
the elements of which are polynomial functions 
of~($z_1,\theta_1,\bar\theta_1$) and~($z_2,\theta_2,\bar\theta_2$). 
We solve this defining relation in the space of lowest weights.
All this is the short summary of our previous paper~\cite{DKK} 
where the reader can find all details.

In Section 5 we reduce the defining system of 
differential equations of the second order for 
the kernel of R-matrix to the equivalent system of the 
first-order differential equations.
Next we demonstrate by direct calculation 
that the obtained integral operator coincides with 
ones obtained in our previous paper. 

Finally, in Section 6 we summarize.
Appendix A contains some technical details about 
the construction of superconformal N-point functions.
In Appendix B we give the list of superintegrals which 
are needed for the calculations of matrix elements.

\section{$s\ell(2)$-invariant $\R$-matrix}
\setcounter{equation}{0}

\subsection{Algebra $s\ell(2)$ and 
lowest weight modules}
\setcounter{equation}{0}
\label{sec:sl2}

The algebra $s\ell(2)$ has three 
generators~$\vec S = (S^1,S^2,S^3)$ with 
standard commutation relations: 
$$
[S^j,S^k]= i \epsilon^{jkp}S^p
$$
We shall use the generators $S,S^{\pm}$:
$$
S^{\pm} = S^1 \pm i S^2\ ,\ S = S^3 
\ ;\ [S,S^{\pm}] =\pm S^{\pm}\ ,\  [S^{+},S^{-}] = 2 S. 
$$
which have the following form in the 
fundamental representation 
$$
S^i = \half \sigma^i
\ ;\ \sigma_{1}
= \left (\begin{array}{cc}
0 & 1 \\
1 & 0 \end{array} \right )
\ ;\  \sigma_{2}
= \left (\begin{array}{cc}
0 & -i \\
i & 0 \end{array} \right )
\ ;\  \sigma_{3}
= \left (\begin{array}{cc}
1 & 0 \\
0 & -1 \end{array} \right ),
$$
$$
S^{+}
= \left (\begin{array}{cc}
0 & 1 \\
0 & 0 \end{array} \right )
\ ;\  S^{-}
= \left (\begin{array}{cc}
0 & 0 \\
1 & 0 \end{array} \right )
\ ;\  S = \half 
\left (\begin{array}{cc}
1 & 0 \\
0 & -1 \end{array} \right ).
$$ 
We represent the generators as first order differential operators,
acting on the space of polynomials $\Phi(z)$:  
\be
S^{-} = -\partial\ ;\ S^{+}= z^2\partial + 2\ell z
\ ;\ S= z\partial +\ell.
\label{gen}
\ee
Below we shall use the global form 
of $s\ell(2)$-transformations
\be
e^{\lambda S^{-}}\Phi(z)=
\Phi(z-\lambda)  
\ ;\ e^{\lambda S^{+}}\Phi(z)=
\frac{1}{(1-\lambda z)^{2\ell}}
\Phi\left(\frac{z}{1-\lambda z}\right)
\label{glob}
\ee
The lowest weight $s\ell(2)$-module 
$V_{\ell}$ is built on the lowest weight 
vector $\psi$ obeying:
$$
S^{-}\psi = 0\ ;\ S \psi = \ell\psi
$$
The center of the enveloping algebra of the algebra $s\ell(2)$ 
is generated by the Casimir operator~$C_2$ and 
the module is characterised uniquely by the action of 
the Casimir operators on its elements:
\be
C_2 = S^2-S+S^{+}S^{-} 
\ ;\ C_2 v = \ell(\ell-1) v.
\label{C_2}
\ee 
It is a vector space spanned by the following basis
$
V_k = (S^{+})^k \psi 
\ ,\  k\in \Z_{+}.
$
We shall use the above realization of the 
$s\ell(2)$-generators as the differential 
operators of first order acting on the
infinite-dimensional(for generic $\ell$) space 
$V_{\ell}$ of polynomials~$\Phi(z)$ of variable $z$.
The coherent state $e^{\lambda S^{+}}\psi$ for the 
lowest weight $\psi = 1$ is the generating function, 
the power expansion in $\lambda$ of which 
produces the basis: 
\be
e^{\lambda S^{+}} 1 = (1-\lambda z)^{-2\ell}
\ ;\  V_{k}=(2\ell)_k z^k  
\label{V}
\ee
There exist some special values of $\ell$:
$\ell = - n\ ;\ n \in \half \Z_+$ , for which 
the module $V_{\ell}$ becomes a 
finite-dimensional vector space.
Indeed it is evident from~(\ref{V}) that all basis vectors 
are equal zero for $k\geq n+1$.
Let us introduce the special notation 
for the fundamental $s\ell(2)$-module: 
$V_{-\half}~\equiv~V_{f}$.\\
The tensor product of two $s\ell(2)$-modules has 
the well known direct sum decomposition:
\be
V_{\ell_1}\otimes V_{\ell_2} = 
\sum_{n=0}^{\infty} V_{\ell_1+\ell_2+n} 
\label{sum}
\ee
For the proof of~(\ref{sum}) in the generic situation  
one has to determine all possible lowest weight vectors 
appearing in the tensor product~$V_{\ell_1}\otimes V_{\ell_2}$.
In the realization on functions of $z$ the space 
$V_{\ell_1}\otimes V_{\ell_2}$ is isomorphic 
to the space of polynomials in two 
variables $\psi(z_1,z_2)$. 
The $s\ell(2)$-generators acting on the
$V_{\ell_1}\otimes V_{\ell_2}$ are just the sums 
of corresponding generators acting in $V_{\ell_i}$.
The lowest weight vectors of the irreducible 
representations in the decomposition of 
$V_{\ell_1}\otimes V_{\ell_2}$ are defined 
as the solutions of the equation:
\be
S^{-}\Psi = 0 
\Rightarrow
\Psi(z_1,z_2) = 
\Psi\left( z_1-z_2 \right)
\label{lwf}
\ee
Now, the lowest weight vectors in the decomposition of 
the tensor product are constructed from functions~(\ref{lwf}) 
being eigenfunctions of generator $S$.
The eigenfunctions of the operator 
$S$ are homogeneous polynomials of degree~$n\ ,\ n \in \Z_+$ 
and finally we obtain that all lowest weights in the space 
$V_{\ell_1}\otimes V_{\ell_2}$ are:
\be
\Psi_n\equiv \left(z_1-z_2\right)^{n}
\ee

\subsection{Yang-Baxter equation and 
general operator $\R_{\ell_1\ell_2}(u)$}
\setcounter{equation}{0}

Let $V_{\ell_i}\ ;\ i=1,2,3$ be three lowest 
weight $s\ell(2)$-modules. 
Let us consider the three operators $\R_{\ell_i\ell_j}(u)$ 
which are acting in $V_{\ell_i}\otimes V_{\ell_j}$ and obey 
the Yang-Baxter equation in the space 
$V_{\ell_3}\otimes V_{\ell_1}\otimes V_{\ell_2}$:
\be
\R_{\ell_3\ell_1}(u)\R_{\ell_3\ell_2}(v)
\R_{\ell_1\ell_2}(v-u)=
\R_{\ell_1\ell_2}(v-u)\R_{\ell_3\ell_2}(v)
\R_{\ell_3\ell_1}(u)
\label{YB}
\ee 
To obtain the defining relation for the general 
$\R$-operator we consider the special case 
$\ell_1=f$ in~(\ref{YB}) 
\be
\R_{f\ell_1}(u)\R_{f\ell_2}(v)
\R_{\ell_1\ell_2}(v-u)=
\R_{\ell_1\ell_2}(v-u)\R_{f\ell_2}(v)
\R_{f\ell_1}(u).
\label{m}
\ee
We can choose the matrix realization in 
$V_{f}$ and use the operator
($\vec S$ are generators acting in $V_{\ell}$) 
\be
\R_{f \ell}(u) = u +\frac{1}{2}+
\vec\sigma \vec S =
\left (\begin{array}{cc}
u+\half + S & S^{-} \\
S^{+} & u+\half -S \end{array} \right ),
\label{Rfl}
\ee
which is the well known solution of 
the Yang-Baxter equation:
$$
\R_{f_1 f_2}(u-v)\R_{f_1 \ell_3}(u)
\R_{f_2 \ell_3}(v)=
\R_{f_2 \ell_3}(v)\R_{f_1 \ell_3}(u)
\R_{f_1 f_2}(u-v)
\ ;\  \R_{f_1 f_2}(u)= 1 + u\cdot\P_{1 2},
$$
where $\P_{1 2}$ is the permutation.
The operators $\R_{f \ell_1}$,$\R_{f \ell_2}$ 
are linear functions of the spectral parameter $u$ 
$$
\R_{f \ell_i}(u) = u+\half + \R_{i}
\ ;\ \R_{i}=
\left (\begin{array}{cc}
S_i & S^{-}_i \\
S^{+}_i & -S_i
\end{array} \right)
\ ;\ i=2,3.
$$
Now the general R-matrix $\R_{\ell_1\ell_2}(u)$ 
acting in the tensor product $V_{\ell_1}\otimes V_{\ell_2}$
of arbitrary modules, is fixed by the condition~(\ref{m}).
After separation of $v+u$ and $v-u$ dependence we 
obtain two equations($v-u\to u$):
\be
(\R_1+ \R_2)\R_{\ell_1\ell_2}(u) 
= \R_{\ell_1\ell_2}(u)(\R_1+ \R_2)
\label{invar}
\ee
\be
\left(\frac{u}{2}\left(\R_1-\R_2\right)+
\R_1\R_2\right)\R_{\ell_1\ell_2}(u) =
\R_{\ell_1\ell_2}(u)
\left(\frac{u}{2}\left(\R_1-\R_2\right)+\R_2\R_1\right)
\label{info}
\ee
The first equation~(\ref{invar}) expresses the 
fact that $\R_{\ell_1\ell_2}(u)$ has to be invariant 
with respect to the action of $s\ell(2)$-algebra 
$$
\left[\vec S, \R_{\ell_1\ell_2}(u)\right] = 0
\ ;\ \vec S =\vec S_1+\vec S_2  
$$
and the second equation is the wanted 
defining relation for the operator $\R_{\ell_1\ell_2}(u)$.
In the matrix form the defining relation reads as follows:
\be
K_{AB} \R_{\ell_1\ell_2}(u) = 
\R_{\ell_1\ell_2}(u)\bar{K}_{AB}
\ ;\  A,B = 1,2
\label{infom}
\ee
where:
$$
K = \left (\begin{array}{cc}
K_{11} & K_{12} \\
K_{21} & K_{22} 
\end{array} \right )\equiv \frac{u}{2}(\R_1-\R_2)+\R_1\R_2
\ ;\  \bar{K}\equiv \frac{u}{2}(\R_1-\R_2)+\R_2\R_1.
$$
Due to $s\ell(2)$-invariance there is only one 
independent equation so that the system of 
equations~(\ref{invar}),~(\ref{info}) for R-matrix 
is equivalent to the system 
\be
K_{12} \R_{\ell_1\ell_2}(u) = 
\R_{\ell_1\ell_2}(u)\bar{K}_{12}
\ ;\ \left[\vec S, \R_{\ell_1\ell_2}(u)\right] = 0
\label{12}
\ee
Indeed, the operators
$\R_{f \ell_1}(u)\R_{f \ell_2}(v)$
and 
$\R_{f \ell_2}(v)\R_{f \ell_1}(u)$ are 
$s\ell(2)$-invariant by construction
$$
\left[\vec S +\half\vec\sigma,
\R_{f \ell_1}(u)\R_{f \ell_2}(v)\right] = 0
\ ;\ \left[\vec S +\half\vec\sigma,
\R_{f \ell_2}(v)\R_{f \ell_1}(u)\right] = 0
$$
so that we obtain the commutation relations between
$s\ell(2)$-generators and $K$-operators  
$$
\left[\vec S, K \right] = -\half \left[\vec \sigma, K \right]
\ ;\ \left[\vec S, \bar K \right] = 
-\half \left[\vec \sigma, \bar K \right].
$$ 
For example  
$$
\left[S^{+}, K_{12} \right] =  K_{11} - K_{22}
\ ,\ \left[S^{+}, K_{11} \right] = - K_{21}
\ ,\ \left[S^{+}, K_{22} \right] =  K_{21}
\ ,\ \left[S^{+}, K_{21} \right] = 0
$$
and it is possible to obtain the full system of 
equations~(\ref{infom}) starting from 
the $K_{12}$-equation in~(\ref{12}).

\subsection{Eigenvalues of operator $\R_{\ell_1\ell_2}(u)$}
\setcounter{equation}{0}

The $s\ell(2)$-invariance of the operator 
$\R_{\ell_1\ell_2}(u)$ allows to simplify the 
eigenvalue problem.
Due to $s\ell(2)$-invariance 
any eigenspace of the operator $\R_{\ell_1\ell_2}$
is a lowest weight $s\ell(2)$-module 
generated by some lowest weight eigenvector. 
Therefore without loss of generality we can 
solve the defining relation~(\ref{info}) in the 
space of lowest weights.
Let us consider in more details the structure of 
eigenspace of the $s\ell(2)$-invariant operator 
acting on the tensor product~$V_{\ell_1}\otimes V_{\ell_2}$.
As we have seen from the direct sum decomposition~(\ref{sum})
for any fixed $n$ the space of lowest weight 
vectors is one-dimensional and therefore the 
operator $\R_{\ell_1\ell_2}$
is diagonal on lowest weight vectors 
$\Psi_{n}$:
\be
\R_{\ell_1\ell_2}(u) \Psi_{n} = R_n \Psi_{n} 
\label{R}
\ee
The relation~(\ref{12}) leads to a simple 
recurrence relation for the eigenvalues $R_{n}$. 
The operator $K_{12}$ commutes with the lowering 
generator $S^{-}$ and therefore $K_{12}$ maps lowest 
weight vectors to lowest weight vector and decrease 
its power by one:  
$$
K_{12}\Psi_{n}=\alpha_{n}(u)\Psi_{n-1}
\ ;\ \alpha_{n}(u) = n(-u+\ell_1+\ell_2+n-1),
$$ 
where the concrete form of the function $\alpha_{n}(u)$ 
is derived by the explicit calculation using
$$
K_{12} = -(z_1-z_2)\dd_1\dd_2 -
\left(\frac{u}{2}-\ell_2\right)\dd_1+
\left(\frac{u}{2}-\ell_1\right)\dd_2
\ ; \  \Psi_{n} = (z_1-z_2)^n.
$$
Under the permutation:
$$
\ell_1,z_1 \leftrightarrow \ell_2,z_2
\ ;\ u \leftrightarrow -u , 
$$
the operators $K_{AB}$ and lowest weights transform as follows: 
$$ 
K_{AB}\leftrightarrow\bar{K}_{AB}
\ ;\  \Psi_{n} \leftrightarrow (-1)^{n+1}\Psi_{n}
$$ 
so that the action of operator $\bar{K}_{12}$ on lowest weight 
vectors can be obtained from formulae for $K_{12}$ 
by the formal substitution $\ell_1,z_1\leftrightarrow \ell_2,z_2$ and 
$u \leftrightarrow -u$:
$$
\bar{K}_{12}\Psi_{n}=-\alpha_{n}(-u)\Psi_{n-1}.
$$
We project the operator equation 
$K_{12}\R =\R\bar{K}_{12}$ onto the lowest weight 
vectors $\Psi_{n}$:
$$
K_{12}\R \Psi_{n} = \R \bar{K}_{12}\Psi_{n}
\Longrightarrow
\alpha_{n}(u) R_{n} \Psi_{n-1} =
-R_{n-1}\alpha_{n}(-u)\Psi_{n-1}
$$
which results in the recurrent relation:
$$
\alpha_{n}(u) R_{n}=
-\alpha_{n}(-u)R_{n-1}
$$
with the general solution~\cite{F,KRS,KS,H}:
\be
R_{n}(u)=(-1)^{n}
\frac{\Gamma\left(-u+\ell_1+\ell_2\right)}
{\Gamma\left(u+\ell_1+\ell_2\right)}
\frac{\Gamma\left(u+\ell_1+\ell_2+n\right)}
{\Gamma\left(-u+\ell_1+\ell_2+n\right)}
\label{eig}
\ee 
where we fixed the overall normalisation of R-operator 
in such a way that $\R: 1 \mapsto 1$.

\section{The operator $\R_{\ell_1\ell_2}(u)$ 
as integral operator}
\setcounter{equation}{0}

It is useful to derive the explicit representation 
of the $\R$-matrix as integral operator acting on polynomials 
$\psi(z_1,z_2)\in V_{\ell_1}\otimes V_{\ell_2}$
\be
\left[\R_{\ell_1\ell_2}(u)\psi\right](z_1,z_2)
= \int\int \rmd z_3 \rmd z_4 
\RR(z_1,z_2|z_3,z_4)\psi(z_3,z_4)
\label{def}
\ee
We suppose that it is possible to integrate by parts so that 
the equations~(\ref{invar}),~(\ref{info}) 
result in the set of differential equations for 
the kernel $\RR(\vec z) = \RR(z_1,z_2|z_3,z_4)$
\be
(\R_1+ \R_2+\R_3+ \R_4)\RR(\vec z) = 0
\label{invar1}
\ee
\be
\left(\frac{u}{2}\left(\R_1-\R_2\right)+
\R_1\R_2\right) \RR(\vec z) =
\left(\frac{u}{2}\left(\R_4-\R_3\right)+\R_4\R_3\right)
\RR(\vec z)
\label{info1}
\ee
where the matrices $\R_i\ ,\ i=3,4$ arise after integration by parts
$$
\R_i =\left (\begin{array}{cc}
S_i & S^{-}_i \\
S^{+}_i & -S_i
\end{array} \right )= 
\left (\begin{array}{cc}
z_i\dd_i+\ell_i & -\dd_i \\
z^2_i\dd_i+2\ell_i z_i & -z_i\dd_i-\ell_i
\end{array} \right )
\ ;\ \ell_3 \equiv 1-\ell_1
\ ,\ \ell_4 \equiv 1-\ell_2 
$$
First we shall solve this system of differential equations 
and then specify selfconsistently the contours of integration. 
The main system of equations can be simplified drastically. 

\begin{prop}
The system of differential equations~(\ref{invar1}),~(\ref{info1}) 
of the second order is equivalent to the 
system of differential equations of the first order
\be
\DD_1\RR = 0\ , \ \DD_2\RR = 0\ , \ \DD_3\RR = 0
\ , \ \DD_4\RR = 0.
\label{sys2}
\ee
where operators $\DD_i$ are defined in a following way
\be
\DD_1 \equiv\dd_1 +\frac{u+\ell_1+\ell_4}{z_{14}}-
\frac{u+\ell_4-\ell_1}{z_{12}}
\ ;\ \DD_2 \equiv \dd_2 +\frac{u+\ell_2+\ell_3}{z_{23}}+
\frac{u+\ell_3-\ell_2}{z_{12}}
\label{oper}
\ee
$$
\DD_3 \equiv \dd_3 -\frac{u+\ell_2+\ell_3}{z_{23}}-
\frac{u+\ell_2-\ell_3}{z_{34}}
\ ;\ \DD_4 \equiv \dd_4 -\frac{u+\ell_1+\ell_4}{z_{14}}+
\frac{u+\ell_1-\ell_4}{z_{34}}. 
$$
The $s\ell(2)$-generators are expressed in terms of $\DD_i$ as follows:
$$
\S^{-} = -\sum_{i=1}^4 \DD_i
\ ;\ \S = \sum_{i=1}^4 z_i\DD_i
\ ;\ \S^{+} = \sum_{i=1}^4 z^2_i\DD_i
\ ;\ \vec{\S} = \sum_{i=1}^4 \vec S_i
$$ 
The common solution of the system~(\ref{sys2}) 
up to overall normalisation is 
$$
\RR(\vec z) \sim 
z_{12}^{u-\ell_1+\ell_4}
z_{14}^{-u-\ell_1-\ell_4}
z_{32}^{-u-\ell_2-\ell_3}
z_{34}^{u+\ell_2-\ell_3}
\ ; \ \ell_3=1-\ell_1\ , \ \ell_4=1-\ell_2.  
$$
\end{prop}

\subsection{Differential equations}

There exists the simple and elegant 
way to derive the equation of the first order for the 
R-matrix which is especially useful for the 
supersymmetric generalisation.
The similar trick was used in~\cite{L,FadK,Kor,KK} for the 
derivation of integrals of motion for 
$XXX$ spin chains.
The matrix $\R$ has very special structure
$$
\R = \left (\begin{array}{cc}
z & -1 \\
z^2 & -z
\end{array} \right )\dd + 
\left (\begin{array}{cc}
\ell & 0 \\
2\ell z & -\ell
\end{array} \right )
\ ;\ 
\left (\begin{array}{cc}
z & -1 \\
z^2 & -z
\end{array} \right )=
\left(\begin{array}{c}
1 \\ z   
\end{array} \right ) 
\left(\begin{array}{cc}
z & -1 
\end{array} \right )
$$
so that the matrix
$$
\R f(z)= 
\left (\begin{array}{cc}
z & -1 \\
z^2 & -z
\end{array} \right ) f^{\prime}(z) + 
\left (\begin{array}{cc}
\ell & 0 \\
2\ell z & -\ell
\end{array} \right ) f(z)
$$
has two evident eigenvectors 
$$
\left(\begin{array}{cc}
z & -1 
\end{array} \right ) \R f(z) =
-\ell 
\left(\begin{array}{cc}
z & -1 
\end{array} \right )f(z)
\ ;\ \R f(z)
\left(\begin{array}{c}
1 \\ z   
\end{array} \right )=
\ell 
\left(\begin{array}{c}
1 \\ z   
\end{array} \right )f(z)
$$
where $f(z)$ is an arbitrary function.
The derivative $f^{\prime}(z)$ disappears from the eigenvalues 
and we shall use this property to derive the 
first order differential equation for the kernel $\RR(\vec z)$.
Let us multiply the equations~(\ref{invar}),~(\ref{info}) 
to the vector 
$
\left(\begin{array}{cc}
z_4 & -1 
\end{array} \right )
$ from the left and to the vector 
$
\left(\begin{array}{c}
1 \\ z_2   
\end{array} \right )
$ 
from the right. 
Then we obtain the system of two 
first-order differential equations 
for the function $\RR(\vec z)$ 
$$
\D_1 \RR + \D_3 \RR = (\ell_2-\ell_4)z_{24} \RR
\ ;\ \left(\frac{u}{2}+\ell_2\right)\D_1 \RR +
\left(\frac{u}{2}+\ell_4\right)\D_3 \RR = 
-\frac{u}{2}(\ell_2-\ell_4)z_{24} \RR 
$$
which is equivalent to the system
\be
\D_1  \RR  = -\left(u+\ell_4\right) z_{24}  \RR 
\ ;\ \D_3  \RR  = \left( u+\ell_2\right) z_{24}  \RR.
\label{sys}
\ee
We use the notation
$$
\D_k \equiv \left(\begin{array}{cc}
z_4 & -1 
\end{array} \right )
\R_k
\left(\begin{array}{c}
1 \\ z_2   
\end{array} \right ) = -z_{k2}z_{k4}\dd_k - \ell_k(z_{k2}+z_{k4})
\ ;\ k=1,3
$$
for the arising differential operators.
Let us multiply the equations~(\ref{invar}),~(\ref{info}) 
to the vector 
$
\left(\begin{array}{cc}
z_1 & -1 
\end{array} \right )
$ from the left and to the vector 
$
\left(\begin{array}{c}
1 \\ z_3   
\end{array} \right )
$ 
from the right. 
In the same manner we obtain the mirror 
system of equations:
\be
\D_2  \RR  = -\left(u+\ell_3\right) z_{13}  \RR 
\ ;\ \D_4  \RR  = \left( u+\ell_1\right) z_{13}  \RR,
\label{sys1}
\ee
where
$$
\D_k \equiv \left(\begin{array}{cc}
z_1 & -1 
\end{array} \right )
\R_k
\left(\begin{array}{c}
1 \\ z_3   
\end{array} \right ) = -z_{k1}z_{k3}\dd_k - \ell_k(z_{k1}+z_{k3})
\ ;\ k=2,4.
$$
Finally, as consequence of the main system of 
differential equations of the second order we obtain 
the system of differential 
equations~(\ref{sys}),~(\ref{sys1}) of the first order.
The eqs.~(\ref{sys}),~(\ref{sys1}) are equivalent to 
$$
\DD_1\RR = 0\ , \ \DD_2\RR = 0\ , \ \DD_3\RR = 0
\ , \ \DD_4\RR = 0.
$$
where the differential operators~$\DD_i$ 
are defined in~(\ref{oper}).\\  
The next step is to prove the reverse, 
that the system of equations~(\ref{invar1}),~(\ref{info1}) 
is the consequence of the system~(\ref{sys2}).
It is easy to rewrite the $s\ell(2)$-generators and 
the equations of $s\ell(2)$-invariance~(\ref{invar1}) in terms of 
the introduced operators $\DD_i$:
$$
\S^{-} = \sum_{i=1}^4 S^{-}_i = -\sum_{i=1}^4 \DD_i
\ ;\ \S = \sum_{i=1}^4 S^{-}_i = \sum_{i=1}^4 z_i\DD_i
\ ;\ \S^{+} = \sum_{i=1}^4 S^{+}_i = \sum_{i=1}^4 z^2_i\DD_i.
$$ 
$$
\S^{-}\RR =\sum_{i=1}^4 \DD_i \RR =0 
\ ;\ \S\RR =\sum_{i=1}^4 z_i\DD_i\RR = 0
\ ;\ \S^{+}\RR =\sum_{i=1}^4 z^2_i\DD_i = 0
$$
As explained above due to $s\ell(2)$-invariance we can 
restrict to one $K_{12}$-equation in~(\ref{12}).
This equation has the explicit form 
$$
z_{12}\dd_1\dd_2 \RR(\vec z)+
z_{34}\dd_3\dd_4 \RR(\vec z) = 
$$
$$
=-\left(\frac{u}{2}-\ell_2\right)\dd_1 \RR(\vec z) + 
\left(\frac{u}{2}-\ell_1\right)\dd_2 \RR(\vec z) -
\left(\frac{u}{2}-\ell_4\right)\dd_3 \RR(\vec z)+
\left(\frac{u}{2}-\ell_3\right)\dd_4 \RR(\vec z)
$$  
and can be easily rewritten in terms of the operators $\DD_i$
$$
z_{12}\DD_1\DD_2\RR +z_{34}\DD_3\DD_4\RR =
$$
$$
=\left(\frac{u}{2}+\ell_3+
(u+\ell_2+\ell_3)\frac{z_{12}}{z_{23}}\right)\DD_1\RR
-\left(\frac{u}{2}+\ell_4-
(u+\ell_1+\ell_4)\frac{z_{12}}{z_{14}}\right)\DD_2\RR+
$$
$$
+\left(\frac{u}{2}+\ell_1-
(u+\ell_1+\ell_4)\frac{z_{34}}{z_{14}}\right)\DD_3\RR-
\left(\frac{u}{2}+\ell_2+
(u+\ell_2+\ell_3)\frac{z_{34}}{z_{23}}\right)\DD_4\RR.
$$
We see that the system of equations~(\ref{invar1}),~(\ref{info1}) 
is the consequence of the system~(\ref{sys2}) and 
finally, the system of the defining equation~(\ref{12})  
is equivalent to the system~(\ref{sys2}).\\ 
The common solution of the system~(\ref{sys2}) 
up to an overall normalisation is 
$$
\RR(\vec z) \sim 
z_{12}^{u-\ell_1+\ell_4}
z_{14}^{-u-\ell_1-\ell_4}
z_{32}^{-u-\ell_2-\ell_3}
z_{34}^{u+\ell_2-\ell_3}
\ ; \ \ell_3=1-\ell_1\ , \ \ell_4=1-\ell_2.  
$$
The operators $\DD_i$ are obtained by the simple 
similarity transformation from the derivatives~$\dd_i$
$$
\DD_1 =z_{12}^{u+\ell_4-\ell_1}z_{14}^{-u-\ell_1-\ell_4} 
\dd_1 z_{12}^{-u-\ell_4+\ell_1}z_{14}^{u+\ell_1+\ell_4}
\ ,\ \DD_2 = z_{21}^{u+\ell_3-\ell_2}z_{23}^{-u-\ell_2-\ell_3}
\dd_2 z_{21}^{-u-\ell_3+\ell_2}z_{23}^{u+\ell_2+\ell_3}
$$
$$
\DD_3 = z_{32}^{-u-\ell_2-\ell_3}z_{34}^{u+\ell_2-\ell_3}
\dd_3 z_{32}^{u+\ell_2+\ell_3}z_{34}^{-u-\ell_2+\ell_3}
\ ,\ \DD_4 = z_{41}^{-u-\ell_1-\ell_4}z_{43}^{u+\ell_1-\ell_4}
\dd_4 z_{41}^{u+\ell_1+\ell_4}z_{43}^{-u-\ell_1+\ell_4}. 
$$
so that by using the ansatz 
$$
\RR(\vec z) \sim 
z_{12}^{u-\ell_1+\ell_4}
z_{14}^{-u-\ell_1-\ell_4}
z_{32}^{-u-\ell_2-\ell_3}
z_{34}^{u+\ell_2-\ell_3}\cdot R(\vec z)
$$
the system of equations~$\DD_i\RR(\vec z) = 0$ 
can be reduced to the system~$\dd_i R(\vec z) = 0$   
with the evident solution~$R(\vec z) = const$.

\subsection{The integral representation.}

It remains to fix the contours of integrations.
The $\R$-operator acts on the space of polynomials 
$\psi(z_1,z_2)$ which can be of arbitrary large degree.
The requirement of convergence allows compact contours 
only and there remains the one possibility: 
\be
\left[\R_{\ell_1\ell_2}(u)\psi\right](z_1,z_2)
= \rho \int_{z_1}^{z_2}\rmd z_3 
\int_{z_1}^{z_3}\rmd z_4 
z_{21}^{1+u-\ell_1-\ell_2}
z_{41}^{\ell_2-\ell_1-u-1}
z_{23}^{\ell_1-\ell_2-u-1}
z_{34}^{\ell_1+\ell_2+u-1}\psi(z_3,z_4)
\label{R1}
\ee  
where the normalisation factor $\rho$ 
is chosen so that $\R: 1\mapsto 1$,
\be
\rho = 
\frac{1}{\Gamma(\ell_1-\ell_2-u)\Gamma(\ell_2-\ell_1-u)}
\frac{\Gamma(\ell_1+\ell_2-u)}{\Gamma(\ell_1+\ell_2+u)}.
\label{norm} 
\ee
It is easy to check~$a\ posteriori$ that 
the obtained integral operator is $s\ell(2)$-invariant 
and has the eigenvalues~(\ref{eig}).
The eigenvalues can be calculated by using twice the general formula
$$
\int_{z_1}^{z_3}\rmd z_4 
z_{34}^{a-1}z_{41}^{b-1} =
z_{31}^{a+b-1}
\frac{\Gamma(a)\Gamma(b)}{\Gamma(a+b)}.
$$

\section{The supersymmetric extension:
$s\ell(2|1)$-invariant $\R$-matrix}
\setcounter{equation}{0}

\subsection{The algebra $s\ell(2|1)$ and 
lowest weights modules}
\setcounter{equation}{0}

Here we collect the main formulas about 
algebra~$s\ell(2|1)$~(for details see~\cite{SNR,JG,Mar,DIC,DKK}).
We represent the generators as first order differential operators,
acting on the space of polynomials $\Phi(z,\theta,\bar\theta)$:   
\be
S^{-} = -\partial
\ ;\ V^{-} = \partial_{\theta}+\frac{1}{2}\bar\theta\partial
\ ;\ W^{-} = \partial_{\bar\theta}+\frac{1}{2}\theta\partial,
\label{gen-}
\ee
$$
V^{+}= -\left[z\partial_{\theta} +\frac{1}{2}\bar\theta z\partial+
\frac{1}{2}\bar\theta\theta\partial_{\theta}\right]-(\ell-b)\bar\theta
\ ;\ W^{+}= -\left[z\partial_{\bar\theta} +\frac{1}{2}\theta z\partial+
\frac{1}{2}\theta\bar\theta\partial_{\bar\theta}\right]-(\ell+b)\theta ,
$$
\be
S^{+}= z^2\partial + z\theta\partial_{\theta}+ 
z\bar\theta\partial_{\bar\theta} + 2\ell z 
-b\theta\bar\theta , 
\label{gen+}
\ee 
$$
S= z\partial +\frac{1}{2}\theta\partial_{\theta}+
\frac{1}{2}\bar\theta\partial_{\bar\theta} +\ell
\ ;\  B= \frac{1}{2}\bar\theta\partial_{\bar\theta} -
\frac{1}{2}\theta\partial_{\theta}+b.
$$
The lowest weight $s\ell(2|1)$-module 
$V_{\ell,b} = V_{\vec\ell}\ ,\ \vec\ell = (\ell,b)$ 
is built on the lowest weight vector $\psi$ obeying:
$$
V_{-} \psi = 0\ ;\  W^{-}\psi = 0
\ ;\  S^{-}\psi = 0\ ;\ S \psi = \ell\psi
\ ;\  B\psi = b \psi
$$
The module $ V_{\vec\ell}$ is a vector space spanned by the following 
basis~\cite{SNR},~\cite{Mar} with the even vectors
$$
A_k = (S^{+})^k \psi \ ;\  B_k = (S^{+})^{k-1}W^{+}V^{+}\psi
\ ,\  k\in \Z_{+}
$$
and the odd vectors
$$
V_k = (S^{+})^k V^{+}\psi \ ;\  W_k = (S^{+})^{k}W^{+}\psi
\ ,\  k\in \Z_{+}
$$
We shall use the above realization of the 
$s\ell(2|1)$-generators as the differential 
operators of first order acting on the
infinite-dimensional(for generic $\ell$) space 
$V_{\ell,b}$ of polynomials~$\Phi(z,\theta,\bar\theta)$ 
of variables $z,\theta,\bar\theta$.\\
The tensor product of two $s\ell(2|1)$-modules has 
the following direct sum decomposition~\cite{Mar,DKK}:
\be
V_{\ell_1,b_1}\otimes V_{\ell_2,b_2} = V_{\ell,b}+
2\sum_{n=1}^{\infty} V_{\ell+n,b}+
\sum_{n=0}^{\infty} V_{\ell+n+\half,b-\half}+
\sum_{n=0}^{\infty} V_{\ell+n+\half,b+\half}  
\ ;\ \ell_i\ne\pm b_i 
\label{Ssum}
\ee
$$
\ell = \ell_1+\ell_2\ ;\ b = b_1+b_2
$$
Note that this formula is applicable in the 
generic situation $\ell_i\ne\pm b_i$.
All possible lowest weight vectors 
appearing in the tensor 
product~$V_{\ell_1,b_1}\otimes V_{\ell_2,b_2}$ 
are divided in two sets, the even lowest weights:
\be
\Phi^{\pm}_n\equiv 
\left(Z_{12}\pm\half\theta_{12}\bar\theta_{12}\right)^{n}
\ ;\ D_1^{\pm}\Phi^{\pm}_n = 0
\ ,\ S \Phi^{\pm}_n = (n +\ell)\Phi^{\pm}_n
\ ,\ B \Phi^{\pm}_n = b \Phi^{\pm}_n
\label{boson}
\ee
and the odd lowest weights: 
\be
\Psi^{-}_n\equiv\theta_{12}Z_{12}^{n}
\ ;\ \Psi^{+}_n\equiv\bar\theta_{12}Z_{12}^{n}
\ ;\ S \Psi^{\pm}_n = (n +\ell+\half)\Psi^{\pm}_n
\ ,\ B \Psi^{\pm}_n = (b\pm\half)\Psi^{\pm}_n
\label{fermion}
\ee
where 
$$
Z_{ij}\equiv 
z_i-z_j+\frac{1}{2}(\bar\theta_i\theta_j+\theta_i\bar\theta_j)
\ ;\ \theta_{ij}\equiv \theta_i - \theta_j
\ ;\ \bar\theta_{ij}\equiv \bar\theta_i - \bar\theta_j.
$$

\subsection{Yang-Baxter equation and 
general operator $\R_{\vec\ell_1\vec\ell_2}(u)$}
\setcounter{equation}{0}

Let $V_{\ell_i,b_i}\ ;\ i=1,2,3$ be three lowest 
weight $s\ell(2|1)$-modules. 
We shall use the short-hand notation:
$$
\vec f =\left(-\half,-\half\right)\ ;\ \vec\ell=(\ell,b)
\ ;\  V_{\vec\ell}=V_{\ell,b}.
$$
Let us consider the three operators $\R_{\vec\ell_i\vec\ell_j}(u)$ 
which are acting in $V_{\vec\ell_i}\otimes V_{\vec\ell_j}$ and obey 
the Yang-Baxter equation in the space 
$V_{\vec\ell_3}\otimes V_{\vec\ell_1}\otimes V_{\vec\ell_2}$~\cite{KS}:
\be
\R_{\vec\ell_3\vec\ell_1}(u)\R_{\vec\ell_3\vec\ell_2}(v)
\R_{\vec\ell_1\vec\ell_2}(v-u)=
\R_{\vec\ell_1\vec\ell_2}(v-u)\R_{\vec\ell_3\vec\ell_2}(v)
\R_{\vec\ell_3\vec\ell_1}(u)
\label{sYB}
\ee 
To obtain the defining relation for the general 
$\R$-operator we consider the special case 
${\vec\ell}_3=\vec f$ in~(\ref{sYB}).  
Then one can choose the matrix realization~\cite{K,FPT,DKK} 
in $V_{\vec\ell_1}$ 
$$
\R_{\vec f\vec\ell}(u)  =
\left (\begin{array}{ccc}
u+S+B & - W^{-} &  S^{-} \\
 V^{+} & u+2 B  &  V^{-} \\
 S^{+} &  W^{+}  &  u+B-S 
\end{array} \right )
$$
and the operators 
$\R_{\vec f\vec\ell_1}$,$\R_{\vec f\vec\ell_2}$ 
are linear functions of spectral parameter $u$ 
in this particular case
$$
\R_{\vec f\vec\ell_i}(u) = u + \R_{i}
\ ;\ \R_{i}=
\left (\begin{array}{ccc}
S_i+B_i & -W^{-}_i & S^{-}_i \\
V^{+}_i & 2 B_i  & V^{-}_i \\
S^{+}_i & W^{+}_i & B_i-S_i
\end{array} \right )
\ ;\ i=1,2
$$
Now the general R-matrix $\R_{\vec\ell_1\vec\ell_2}(u)$ 
acting in the tensor product $V_{\vec\ell_1}\otimes V_{\vec\ell_2}$
of arbitrary modules, is fixed by the condition
\be
\R_{\vec f\vec\ell_1}(u)
\R_{\vec f\vec \ell_2}(v)\R_{\vec\ell_1\vec\ell_2}(v-u) = 
\R_{\vec\ell_1\vec\ell_2}(v-u)\R_{\vec f\vec\ell_2}(v)
\R_{\vec f\vec\ell_1}(u)
\label{Sm}
\ee
or equivalently:
\be
(\R_1+ \R_2)\R_{\vec\ell_1\vec\ell_2}(u) =
\R_{\vec\ell_1\vec\ell_2}(u)(\R_1+ \R_2)
\label{sinvar}
\ee
\be
\left(\frac{u}{2}(\R_1-\R_2)+\R_1\R_2\right)
\R_{\vec\ell_1\vec\ell_2}(u) =
\R_{\vec\ell_1\vec\ell_2}(u)
\left(\frac{u}{2}(\R_1-\R_2)+\R_2\R_1\right)
\label{sinfo}
\ee
The first equation~(\ref{invar}) expresses the 
fact that $\R(u)$ has to be invariant 
with respect to the action of $s\ell(2|1)$-algebra and the 
second equation is the defining relation for the 
operator $\R_{\vec\ell_1\vec\ell_2}(u)$.

\subsection{Eigenvalues of operator 
$\R_{\vec\ell_1\vec\ell_2}(u)$}
\setcounter{equation}{0}

Due to $s\ell(2|1)$-invariance 
any eigenspace of the operator $\R_{\vec\ell_1\vec\ell_2}$
is a lowest weight $s\ell(2|1)$-module 
generated by some lowest weight eigenvector. 
As we have seen from direct sum decomposition~(\ref{Ssum})
for every fixed $n$ the space of lowest weight vectors with 
eigenvalue $b$ is two-dimensional and the ones with
eigenvalues $b\pm \half$ are one-dimensional.
Therefore the operator $\R_{\vec\ell_1\vec\ell_2}$
is diagonal on odd lowest weight vectors 
$\Psi^{+}_{n}$ and $\Psi^{-}_{n}$
but acts non-trivially on the two-dimensional subspace of 
even lowest weight vectors spanned on 
$\Phi^{+}_{n}$ and $\Phi^{-}_{n}$.\\
In matrix form we have:
\be
\R_{\vec\ell_1\vec\ell_2}(u)
\left (\begin{array}{c}
\Phi^{+}_{n} \\  \Phi^{-}_{n} \\ \Psi^{+}_{n}\\ 
\Psi^{-}_{n} 
\end{array} \right ) =
\left (\begin{array}{cccc}
A_{n}(u) & B_{n}(u) & 0 & 0 \\
C_{n}(u) & D_{n}(u) & 0 & 0 \\
0 & 0 & F_{n}(u) & 0 \\
0 & 0 & 0 & E_{n}(u) \\
\end{array} \right )
\left (\begin{array}{c}
\Phi^{+}_{n} \\  \Phi^{-}_{n} \\ \Psi^{+}_{n}\\ 
\Psi^{-}_{n} 
\end{array} \right )
\label{ABCD}
\ee
The matrix relation~(\ref{info}) leads to a set of 
recurrence relations for the coefficients $A_{n},...,E_{n}$ 
with the general solution~\cite{DKK}:
\be
A_{n}(u)=(-1)^{n+1}
\frac{\Gamma\left(\u+\ell_n\right)}
{\Gamma\left(-\u+\ell_n+1\right)}
\cdot
\label{solABCD}
\ee
$$
\cdot\frac{
(\ell_1-b_1)(\ell_1+b_1)\left(\u-b_{1}-b_{2}\right)-
\left(\u+b_{1}+b_{2}\right)
\left(\u-b_1-\ell_2\right)\left(\u-b_1+\ell_2\right)}
{(\ell_2-b_2)(\ell_1+b_1)}
$$
$$
B_{n}(u)=(-1)^{n}
\frac{\Gamma\left(\u+\ell_n\right)}
{\Gamma\left(-\u+\ell_n\right)}
\ ;\ C_{n}(u)=(-1)^{n}
\frac{\Gamma\left(\u+\ell_n+1\right)}
{\Gamma\left(-\u+\ell_n+1\right)}
\cdot\frac{(\ell_2+b_2)(\ell_1-b_1)}{(\ell_2-b_2)(\ell_1+b_1)}
$$
$$
D_{n}(u)=(-1)^{n+1}
\frac{\Gamma\left(\u+\ell_n+1\right)}
{\Gamma\left(-\u+\ell_n\right)}
\cdot\frac{\u+b_{2}-b_{1}}{(\ell_2-b_2)(\ell_1+b_1)}
$$
$$
E_{n}(u)=(-1)^{n}
\frac{\Gamma\left(\u+\ell_n+1\right)}
{\Gamma\left(-\u+\ell_n+1\right)}
\cdot\frac{(\u+b_2-\ell_1)(\u+b_2+\ell_1)}
{(\ell_2-b_2)(\ell_1+b_1)}
$$
$$
F_{n}(u)=(-1)^{n}
\frac{\Gamma\left(\u+\ell_n+1\right)}
{\Gamma\left(-\u+\ell_n+1\right)}
\cdot\frac{(\u-b_1-\ell_2)(\u-b_1+\ell_2)}
{(\ell_2-b_2)(\ell_1+b_1)}
$$
where we used the notations:
$$
\ell_n\equiv n + \ell_1+\ell_2
\ ;\ \u\equiv u+ b_2-b_1.
$$
As usual the obtained general solution of the Yang-Baxter 
equation is fixed up to overall normalization.
We choose the normalization such that the R-matrix 
coincides with the permutation operator for 
$u=0$ and $\vec\ell_1=\vec\ell_2$.
Note that the transformation from our basis vectors 
$\Phi^{\pm}_{n}$ to the ones diagonalizing the R-matrix 
depends on $n$ and this does not result in 
simpler formulae.
\footnote{Note that we have admitted some inconsistency 
in notations in our previous work~\cite{DKK}.
We have started from the general Yang-Baxter 
equation
$$
\R_{\vec\ell_1\vec\ell_2}(u-v)\R_{\vec\ell_1\vec \ell_3}(u)
\R_{\vec\ell_2\vec \ell_3}(v)=
\R_{\vec\ell_2\vec \ell_3}(v)\R_{\vec\ell_1\vec \ell_3}(u)
\R_{\vec\ell_1\vec\ell_2}(u-v)
$$
have choosen $\vec \ell_3 = \vec f$ and obtained
$$
\R_{\vec\ell_1\vec\ell_2}(u-v)\R_{\vec\ell_1\vec f}(u)
\R_{\vec\ell_2\vec f}(v)=
\R_{\vec\ell_2\vec f}(v)\R_{\vec\ell_1\vec f}(u)
\R_{\vec\ell_1\vec\ell_2}(u-v)
$$
Then we have ``automatically'' used the ansatz for $\R_{\vec f \vec\ell}$
instead of $\R_{\vec\ell \vec f}$.
As a consequence the results for the matrix 
elements formulated there are referring 
actually to the operator $\R_{\vec\ell_2\vec\ell_1}(u)$
$$ 
\R_{\vec\ell_2\vec\ell_1}(u)\equiv 
\left.\P \R_{\vec\ell_1\vec\ell_2}(u)
\P \right|_{\vec\ell_1\leftrightarrow\vec\ell_2}
\ ;\ \P\psi(Z_1,Z_2) \equiv \psi(Z_2,Z_1).
$$  
In the simplest $s\ell(2)$-case these 
matrices coincide
$
\R_{\ell_1\ell_2}(u) = \R_{\ell_2\ell_1}(u).
$
In the case of superalgebra $s\ell(2|1)$
the situation is unusual because the 
flip-operator $\P$ acts nontrivially 
in the lowest weights basis we have used.  
The correct answer in selfconsistent 
notations is given in~(\ref{solABCD}).}

\section{The operator $\R_{\vec\ell_1\vec\ell_2}(u)$ 
as integral operator}
\setcounter{equation}{0}

Let us go to the derivation of the explicit representation for 
the $\R$-matrix as integral operator acting on polynomials 
$\psi(Z_1,Z_2)\in V_{\vec\ell_1}\otimes V_{\vec\ell_2}$
\be
\left[\R_{\vec\ell_1\vec\ell_2}(u)\psi\right](Z_1,Z_2)
= \int\int \rmd Z_3 \rmd Z_4 
\RR(Z_1,Z_2|Z_3,Z_4)\psi(Z_3,Z_4)
\label{sdef}
\ee
where $Z\equiv\{z,\theta,\bar\theta\}$ and we use the 
standard superintegration
$$
\int \rmd Z \equiv 
\int \rmd z \int \rmd \theta \int \rmd \bar\theta. 
$$  
We suppose that it is possible to integrate by parts so that 
the equations~(\ref{sinvar}),~(\ref{sinfo}) 
give the set of differential equations for 
the kernel $\RR(\vec Z) = \RR(Z_1,Z_2| Z_3, Z_4)$
\be
(\R_1+ \R_2+\R_3+ \R_4)R(\vec  Z) = 0
\label{sinvar1}
\ee
\be
\left(\frac{u}{2}\left(\R_1-\R_2\right)+
\R_1\R_2\right) R(\vec  Z) =
\left(\frac{u}{2}\left(\R_4-\R_3\right)+\R_4\R_3\right)
R(\vec  Z)
\label{sinfo1}
\ee
where the matrices $\R_i\ ,\ i=3,4$ arise after integration by parts
$$
\R_{i}(u)  =
\left (\begin{array}{ccc}
S_i+B_i & - W^{-}_i &  S^{-}_i \\
V^{+}_i & 2 B_i  &  V^{-}_i \\
S^{+}_i &  W^{+}_i  &  B_i-S_i 
\end{array} \right )
\ ;\ \vec\ell_3 \equiv -\vec\ell_1=(-\ell_1,-b_1)
\ ,\ \vec\ell_4 \equiv -\vec\ell_2=(-\ell_2,-b_2)
$$
First we shall solve this system of differential equations 
and then specify selfconsistently the contours of integration.\\

\begin{prop}
The system of differential equations~(\ref{sinvar1}),~(\ref{sinfo1}) 
of second order is equivalent to the 
system of differential equations of first order
\be
\DD_i\RR = 0\ , \ \DD_i^{\pm}\RR = 0\ ;\ i=1,2,3,4.
\label{Ssys2}
\ee
where the operators $\DD_i, \DD_i^{\pm}$ are defined in a following way
$$
\DD_1 \equiv\dd_1-
\frac{\bar\theta_{12}}{Z_{12}}D^{-}_1-
\frac{\theta_{14}}{Z_{14}}D^{+}_1 
+\frac{\u+\ell_1-\ell_2}{Z^{+}_{14}}-
\frac{\u-\ell_1-\ell_2}{Z^{-}_{12}} +
(\u-\ell_2-b_1)\frac{\theta_{14}\bar\theta_{12}}{Z_{14}Z_{12}}
$$
\be
\DD^{-}_1 \equiv D^{-}_1 -
(\u-\ell_1-\ell_2)\frac{\theta_{12}}{Z_{12}}+
(\u-\ell_2-b_1)\frac{\theta_{14}}{Z_{14}}
\label{soper1}
\ee
$$
\DD^{+}_1 \equiv D^{+}_1 +
(\ell_1+b_1)\frac{\bar\theta_{14}}{Z_{14}}+
(\u+b_2-b_1)\frac{Z^{-}_{43}}{Z^{-}_{41}}
\left(\frac{\bar\theta_{43}}{Z_{43}}-
\frac{\bar\theta_{23}}{Z_{23}}\right)
$$
$$
\DD_2 \equiv\dd_2-
\frac{\bar\theta_{23}}{Z_{23}}D^{-}_2-
\frac{\theta_{21}}{Z_{21}}D^{+}_2 
-\frac{\u-\ell_1-\ell_2}{Z^{+}_{21}}+
\frac{\u-\ell_1+\ell_2}{Z^{-}_{23}} -
(\u-\ell_1+b_2)\frac{\theta_{21}\bar\theta_{23}}{Z_{21}Z_{23}}
$$
\be
\DD^{+}_2 \equiv D^{+}_2 -
(\u-\ell_1-\ell_2)\frac{\bar\theta_{21}}{Z_{21}}+
(\u-\ell_1+b_2)\frac{\bar\theta_{23}}{Z_{23}}
\label{soper2}
\ee
$$
\DD^{-}_2 \equiv D^{-}_2 +
(\ell_2-b_2)\frac{\theta_{23}}{Z_{23}}+
(\u+b_2-b_1)\frac{Z^{+}_{34}}{Z^{+}_{32}}
\left(\frac{\theta_{43}}{Z_{43}}-
\frac{\theta_{14}}{Z_{14}}\right)
$$
$$
\DD_3 \equiv\dd_3-
\frac{\bar\theta_{32}}{Z_{32}}D^{-}_3-
\frac{\theta_{34}}{Z_{34}}D^{+}_3 
-\frac{\u+\ell_1+\ell_2}{Z^{+}_{34}}+
\frac{\u-\ell_1+\ell_2}{Z^{-}_{32}} -
(\u+\ell_2-b_1)\frac{\theta_{34}\bar\theta_{32}}{Z_{34}Z_{32}}
$$
\be
\DD^{+}_3 \equiv D^{+}_3 -
(\u+\ell_1+\ell_2)\frac{\bar\theta_{34}}{Z_{34}}+
(\u+\ell_2-b_1)\frac{\bar\theta_{32}}{Z_{32}}
\label{soper3}
\ee
$$
\DD^{-}_3 \equiv D^{-}_3 -
(\ell_1-b_1)\frac{\theta_{23}}{Z_{23}}+
(\u+b_2-b_1)\frac{Z^{+}_{21}}{Z^{+}_{23}}
\left(\frac{\theta_{12}}{Z_{12}}-
\frac{\theta_{14}}{Z_{14}}\right)
$$
$$
\DD_4 \equiv\dd_4-
\frac{\bar\theta_{43}}{Z_{43}}D^{-}_4-
\frac{\theta_{41}}{Z_{41}}D^{+}_4 
+\frac{\u+\ell_1-\ell_2}{Z^{+}_{41}}-
\frac{\u+\ell_1+\ell_2}{Z^{-}_{43}} +
(\u+\ell_1+b_2)\frac{\theta_{41}\bar\theta_{43}}{Z_{41}Z_{43}}
$$
\be
\DD^{-}_4 \equiv D^{-}_4 -
(\u+\ell_1+\ell_2)\frac{\theta_{43}}{Z_{43}}+
(\u+\ell_1+b_2)\frac{\theta_{41}}{Z_{41}}
\label{soper4}
\ee
$$
\DD^{+}_4 \equiv D^{+}_4 -
(\ell_2+b_2)\frac{\bar\theta_{41}}{Z_{41}}+
(\u+b_2-b_1)\frac{Z^{-}_{12}}{Z^{-}_{14}}
\left(\frac{\bar\theta_{12}}{Z_{12}}-
\frac{\bar\theta_{23}}{Z_{23}}\right)
$$ 
where 
$$
D^{+}_i = -\dd_{\theta_i}+\half\bar\theta_i\dd_i 
\ ;\ D^{-}_i = -\dd_{\bar\theta_i}+\half\theta_i\dd_i
$$
$$
Z^{\pm}_{ik} = Z_{ik}\pm\half \theta_{ik}\bar\theta_{ik}
\ ;\ D^{\pm}_iZ^{\pm}_{ik}=0 
\ ;\ D^{\mp}_kZ^{\pm}_{ik}=0. 
$$ 
This system of twelve equations is invariant with 
respect the following discrete symmetry transformations:
$$
1\leftrightarrow 4
\ ,\ 2 \leftrightarrow 3
\ ;\ (\ell_1,b_1)\leftrightarrow (-\ell_2,-b_2),  
$$ 
$$
1\leftrightarrow 2
\ ,\ 3 \leftrightarrow 4
\ ;\ \theta \leftrightarrow \bar\theta
\ ,\ D^{+}\leftrightarrow D^{-}
\ ;\ (\ell_1,b_1)\leftrightarrow (\ell_2,-b_2).
$$ 
\end{prop}

\subsection{Differential equations}

We use the same trick as in the $s\ell(2)$-case 
and derive the simple first order 
differential equations for the kernel of R-operator.  
The matrix $\R$ has a very special structure
$$
\R = \left(\begin{array}{c}
1 \\ -\bar\theta \\ z+\half\theta\bar\theta   
\end{array} \right ) 
\left(\begin{array}{ccc}
z-\half\theta\bar\theta & -\theta & -1
\end{array} \right )\dd - 
\left(\begin{array}{c}
1 \\ -\bar\theta \\ z+\half\theta\bar\theta   
\end{array} \right ) 
\left(\begin{array}{ccc}
\bar\theta & -1 & 0
\end{array} \right )D^{-}-
$$
$$
-\left(\begin{array}{c}
0 \\ -1 \\ \theta   
\end{array} \right ) 
\left(\begin{array}{ccc}
z-\half\theta\bar\theta & \theta & -1
\end{array} \right )D^{+}+
(\ell+b)
\left (\begin{array}{ccc}
1 & 0 & 0\\
0 & 1 & 0\\
z-\half\theta\bar\theta & -\theta & 0 
\end{array} \right )+
(\ell-b)
\left (\begin{array}{ccc}
0 & 0 & 0\\
-\bar\theta & -1 & 0\\
z+\half\theta\bar\theta & 0 & -1 
\end{array} \right )
$$
where 
$$
D^{+}= -\dd_{\theta}+\half\bar\theta\dd 
\ ;\ D^{-}= -\dd_{\bar\theta}+\half\theta\dd,
$$ 
so that the matrix
$\R f(Z)$, where $f(Z)$ is an even function,
has two evident eigenvectors 
$$
\left(\begin{array}{ccc}
z-\half\theta\bar\theta & -\theta & -1  
\end{array} \right ) \R f =
-(\ell-b) f\cdot \left(\begin{array}{ccc}
z-\half\theta\bar\theta & -\theta & -1  
\end{array} \right )
$$
$$
\R f
\left(\begin{array}{c}
1 \\ -\bar\theta \\ z+\half\theta\bar\theta  
\end{array} \right )=
(\ell+b) f\cdot \left(\begin{array}{c}
1 \\ -\bar\theta \\ z+\half\theta\bar\theta  
\end{array} \right ) 
$$
and all derivatives $\dd f, D^{+}f, D^{-}f$ 
disappear from the relations.
Moreover, there are two other vectors with 
the properties 
$$
\left(\begin{array}{ccc}
\theta & 1 & 0  
\end{array} \right ) \R f =
2b f\cdot\left(\begin{array}{ccc}
\theta & 1 & 0  
\end{array} \right )+
\left(\begin{array}{ccc}
z-\half\theta\bar\theta & \theta & -1  
\end{array} \right ) D^{+}f
$$
$$
\R f
\left(\begin{array}{c}
0 \\ 1 \\ -\theta  
\end{array} \right )=
2b f\cdot
\left(\begin{array}{c}
0 \\ 1 \\ -\theta  
\end{array} \right ) +
\left(\begin{array}{c}
1 \\ -\bar\theta \\ z+\half\theta\bar\theta  
\end{array} \right ) D^{-}f. 
$$
We shall derive the first order differential equations 
for the kernel $\RR(\vec Z)$ by multiplying the 
equations~(\ref{sinvar}),~(\ref{sinfo}) by the appropriate 
vectors from the left and from the right.
Let us start from the three variants for the 
choose of the such vectors. 
First we obtain the system of two first-order 
differential equations for the function $\RR(\vec Z)$ 
$$
\D_1 \RR + \D_3 \RR = (\ell_2-b_2-\ell_4+b_4)Z^{+}_{24}\RR 
$$
$$
\left(\frac{u}{2}+\ell_2+b_2\right)\D_1 \RR +
\left(\frac{u}{2}+\ell_4-b_4\right)\D_3 \RR = 
-\frac{u}{2}(\ell_2-\ell_4+b_2+b_4)Z^{+}_{24} \RR 
$$
which is equivalent to the system
\be
\D_1 \RR  = -\left(u+\ell_4-b_4\right)Z^{+}_{24}\RR 
\ ;\ \D_3  \RR  = \left( u+\ell_2+b_2\right)Z^{+}_{24} \RR.
\label{Ssys1}
\ee
We use the notation 
$$
\D_k \equiv \left(\begin{array}{ccc}
z_4-\half\theta_4\bar\theta_4 & -\theta_4 & -1  
\end{array} \right )
\R_k \left(\begin{array}{c}
1 \\ -\bar\theta_2 \\ z_2+\half\theta_2\bar\theta_2  
\end{array} \right ) = 
$$
$$
= -Z^{-}_{k2}Z^{+}_{k4}\dd_k + Z^{+}_{k4}\bar\theta_{k2}D^{-}_k 
+Z^{-}_{k2}\theta_{k4}D^{+}_k
-(\ell_k+b_k)Z^{+}_{k4}-(\ell_k-b_k)Z^{-}_{k2}+
2b_k\theta_{k4}\bar\theta_{k2}
\ ;\ k=1,3
$$
for the arising differential operators.
Secondly we obtain the system of differential 
equations 
$$
\B_1 \RR + \B_3 \RR = 
(\ell_4-b_4-2b_2)\theta_{24}\RR + Z^{+}_{24}D^{-}_2\RR 
$$
$$
\left(\frac{u}{2} + 2b_2\right)\B_1 \RR +
\left(\frac{u}{2}+\ell_4-b_4\right)\B_3 \RR = 
-\frac{u}{2}(\ell_4-b_4-2b_2)\theta_{24} \RR 
-\frac{u}{2}Z^{+}_{24}D^{-}_2\RR -\D_1 D^{-}_2\RR
$$
which is equivalent to the system 
\be
\B_1 \RR  = \left(u+\ell_4-b_4\right)\theta_{24}\RR 
\ ;\ \B_3 \RR  = -\left(u+2b_2\right)\theta_{24} \RR+ 
Z^{+}_{24}D^{-}_2 \RR,
\label{Ssys21}
\ee
by conditions~(\ref{Ssys1}) and where  
$$
\B_k \equiv \left(\begin{array}{ccc}
z_4-\half\theta_4\bar\theta_4 & -\theta_4 & -1  
\end{array} \right )
\R_k \left(\begin{array}{c}
0 \\ 1 \\ -\theta_2  
\end{array} \right ) = 
$$
$$
= Z^{+}_{k4}\theta_{k2}\dd_k - Z^{+}_{k4} D^{-}_k 
-\theta_{k2}\theta_{k4}D^{+}_k
+(\ell_k+b_k)\theta_{k2}-(\ell_k-b_k)\theta_{24}
\ ;\ k=1,3.
$$
The last system is analogous to the previous ones: 
$$
\C_1 \RR + \C_3 \RR = 
(\ell_2+b_2+2b_4)\bar\theta_{24}\RR + Z^{+}_{24}D^{+}_4\RR 
$$
$$
\left(\frac{u}{2} +\ell_2+b_2\right)\C_1 \RR +
\left(\frac{u}{2}+1-2b_4\right)\C_3 \RR = 
-\frac{u}{2}(\ell_2+b_2+2b_4)\bar\theta_{24} \RR 
-\frac{u}{2}Z^{+}_{24}D^{+}_4\RR +D^{+}_4\D_3 \RR
$$
and is equivalent to the system 
\be
\C_3 \RR  = \left(u+\ell_2+b_2\right)\bar\theta_{24}\RR 
\ ;\ \C_1 \RR  = -\left(u-2b_4\right)\bar\theta_{24} \RR+ 
Z^{+}_{24}D^{+}_4 \RR,
\label{Ssys3}
\ee
by conditions~(\ref{Ssys1}).
The operators $\C_k$ are   
$$
\C_k \equiv \left(\begin{array}{ccc}
\bar\theta_4 & 1 & 0  
\end{array} \right )
\R_k \left(\begin{array}{c}
1 \\ -\bar\theta_2 \\ z_2+\half\theta_2\bar\theta_2  
\end{array} \right ) = 
$$
$$
= -Z^{-}_{k2}\bar\theta_{k4}\dd_k + 
\bar\theta_{k4}\bar\theta_{k2}D^{-}_k 
+Z^{-}_{k2}D^{+}_k-(\ell_k+b_k)\bar\theta_{24}
-(\ell_k-b_k)\bar\theta_{k2}
\ ;\ k=1,3.
$$
The obtained system of the six equations can be 
transformed to the equivalent form.
The $\D_{1,3}$-equations~(\ref{Ssys1}) are 
equivalent to the equations~$\DD_1\RR=0\ ,\ \DD_3\RR=0$.  
The $\B_1$-equation from~(\ref{Ssys21}) can be 
transformed to the equation~$\DD^{-}_1\RR=0$ by 
the condition~$\DD_1\RR=0$.
The $\C_3$-equation from~(\ref{Ssys3}) can be 
transformed to the equation~$\DD^{+}_3\RR=0$ by 
the condition~$\DD_3\RR=0$.  
The remaining $\B_3,\C_1$-equations are 
equivalent to the more involved equations
$$
-Z^{+}_{34}
\left(1-\frac{\theta_{32}\bar\theta_{32}}{Z_{32}}\right)
\left(D^{-}_3 +
(\u+\ell_2-\ell_1)\frac{\theta_{32}}{Z_{32}}-
(\u+\ell_2-b_1)\frac{\theta_{34}}{Z_{34}}\right)\RR = 
Z^{+}_{24}D^{-}_{2}\RR +(\ell_2-b_2)\theta_{24}\RR
$$
$$
Z^{-}_{12}
\left(1+\frac{\theta_{14}\bar\theta_{14}}{Z_{14}}\right)
\left(D^{+}_1 +
(\u+\ell_1-\ell_2)\frac{\bar\theta_{14}}{Z_{14}}-
(\u-\ell_2-b_1)\frac{\bar\theta_{12}}{Z_{12}}\right)\RR = 
Z^{+}_{24}D^{+}_{4}\RR-(\ell_2+b_2)\bar\theta_{24}\RR
$$
by the conditions~$\DD_1\RR=0\ ,\ \DD_3\RR=0$.

It is possible to derive the mirror system of six equations 
in the same manner by choosing appropriate vectors 
for multiplication from the left and from the right. 
In fact, the resulting equations can be obtained from the 
present ones by the symmetry transformation
$$
1\to 4\ ,\ 2 \to 3\ ;\ (\ell_1,b_1)\to (\ell_4,b_4) = (-\ell_2,-b_2)  
\ ,\ (\ell_2,b_2)\to (\ell_3,b_3) = (-\ell_1,-b_1)
$$  
In this way we obtain the 
equations~$\DD_4\RR=0\ ,\ \DD_2\RR=0$ 
and~$\DD^{-}_4\RR=0\ ,\ \DD^{+}_3\RR=0$ and the 
mirror partners of the more involved equations: 
$$
-Z^{+}_{21}
\left(1+\frac{\theta_{32}\bar\theta_{32}}{Z_{32}}\right)
\left(D^{-}_2 +
(\u+\ell_2-\ell_1)\frac{\theta_{32}}{Z_{32}}-
(\u-\ell_1+b_2)\frac{\theta_{21}}{Z_{21}}\right)\RR = 
Z^{+}_{31}D^{-}_{3}\RR -(\ell_1-b_1)\theta_{31}\RR
$$
$$
Z^{-}_{43}
\left(1-\frac{\theta_{14}\bar\theta_{14}}{Z_{14}}\right)
\left(D^{+}_4 +
(\u+\ell_1-\ell_2)\frac{\bar\theta_{14}}{Z_{14}}-
(\u+\ell_1+b_2)\frac{\bar\theta_{43}}{Z_{43}}\right)\RR = 
Z^{+}_{31}D^{+}_{1}\RR+(\ell_1+b_1)\bar\theta_{31}\RR.
$$
The system of four involved equations decomposes into 
two independent systems of two equations.
It can be shown that the system which involves 
the covariant derivatives~$D^{+}_1$ and $D^{+}_4$ 
is equivalent to the system of 
the equations~$\DD^{+}_1\RR=0\ ,\ \DD^{+}_4\RR=0$ and 
the system which involves the covariant 
derivatives~$D^{-}_2$ and $D^{-}_3$ 
is equivalent to the system of 
the equations~$\DD^{-}_2\RR=0\ ,\ \DD^{-}_3\RR=0$. 

The next step is to prove the reverse, 
that the system of equations~(\ref{sinvar1}),~(\ref{sinfo1}) 
is the consequence of the system~(\ref{Ssys2}).
It is easy to rewrite the $s\ell(2|1)$-generators and 
therefore the equations of $s\ell(2)$-invariance~(\ref{sinvar1}) 
in terms of introduced operators $\DD_i, \DD^{\pm}_i$:
$$
\S^{-} = - \sum_{i=1}^4 \DD_i
\ ;\ \V^{-} = \sum_{i=1}^4 \bar\theta_i\DD_i - \DD^{+}_i 
\ ;\ \W^{-} = \sum_{i=1}^4 \theta_i\DD_i - \DD^{-}_i
$$
$$
\S^{+} = \sum_{i=1}^4 z^2_{i}\DD_i -
z_i\theta_i\DD^{+}_i-z_i\bar\theta_i\DD^{-}_i 
\ ;\ \V^{+} = \sum_{i=1}^4 -z_i\bar\theta_i\DD_i +
\left(z_i-\half\theta_i\bar\theta_i\right) \DD^{+}_i 
$$
$$
\W^{+} = \sum_{i=1}^4 -z_i\theta_i\DD_i +
\left(z_i+\half\theta_i\bar\theta_i\right) \DD^{-}_i 
$$
$$
\S = \sum_{i=1}^4 
z_i\DD_i-\half\bar\theta_i \DD^{-}_i-
\half\theta_i \DD^{+}_i
\ ;\ \B = -\half\sum_{i=1}^4 
\theta_i\bar\theta_i\DD_i+
\bar\theta_i \DD^{-}_i-\theta_i \DD^{+}_i
$$
The last step is absolutely analogous to the $s\ell(2)$-case. 
Due to $s\ell(2|1)$-invariance we can use 
one $K_{13}$-equation in~(\ref{sinfo1}) and 
rewrite it in terms of operators~$\DD_i, \DD^{\pm}_i$.
We omit these formulae for simplicity.

\subsection{Solution of the system of differential equations}
 
It is useful to transform the main system of 
equations~(\ref{soper1})-(\ref{soper4}) 
to a simpler homogeneous form.
We shall use the following ansatz for this purpose  
$$
\RR(\vec Z) = 
Z_{12}^{a}Z_{14}^{b}
Z_{32}^{c}Z_{34}^{d}\cdot
\left(1+\frac{\theta_{12}\bar\theta_{12}}{Z_{12}}\right)^{\alpha}
\left(1+\frac{\theta_{14}\bar\theta_{14}}{Z_{14}}\right)^{\beta}
\left(1+\frac{\theta_{32}\bar\theta_{32}}{Z_{32}}\right)^{\gamma}
\left(1+\frac{\theta_{34}\bar\theta_{34}}{Z_{34}}\right)^{\delta}
R(\vec Z)
$$
$$
a\equiv \u-\ell_1-\ell_2
\ ;\  b\equiv -\u-\ell_1+\ell_2
\ ;\  c\equiv -\u+\ell_1-\ell_2
\ ;\  d\equiv \u+\ell_1+\ell_2 
$$
$$
\beta=b_1-\alpha
\ ;\ \gamma = -b_2-\alpha
\ ;\ \delta = b_2-b_1+\alpha. 
$$

\begin{prop}
The system of differential 
equations~(\ref{soper1})-(\ref{soper4}) 
for the function $\RR(\vec Z)$ is transformed 
to the simpler system of differential equations for the 
function $R(\vec Z)$
\be
\DD_i R(\vec Z) = 0\ , \ \DD_i^{\pm} R(\vec Z) = 0\ ;\ i=1,2,3,4.
\label{Sssys2}
\ee
where operators $\DD_i, \DD_i^{\pm}$ are defined in a following way
\be
\DD_1 \equiv\dd_1-
\frac{\bar\theta_{12}}{Z_{12}}D^{-}_1-
\frac{\theta_{14}}{Z_{14}}D^{+}_1 
\ ;\ \DD^{-}_1 \equiv D^{-}_1 -
A\left(\frac{\theta_{12}}{Z_{12}}-
\frac{\theta_{14}}{Z_{14}}\right)
\label{soper11}
\ee
$$
\DD^{+}_1 \equiv D^{+}_1 + 
A\left(\frac{\bar\theta_{12}}{Z_{12}}-
\frac{\bar\theta_{14}}{Z_{14}}\right) +
(A+B)\frac{Z^{-}_{43}}{Z^{-}_{41}}
\left(\frac{\bar\theta_{43}}{Z_{43}}-
\frac{\bar\theta_{23}}{Z_{23}}\right)
$$
$$
\DD_2 \equiv\dd_2-
\frac{\bar\theta_{23}}{Z_{23}}D^{-}_2-
\frac{\theta_{21}}{Z_{21}}D^{+}_2 
\ ;\ \DD^{+}_2 \equiv D^{+}_2 -
A\left(\frac{\bar\theta_{21}}{Z_{21}}-
\frac{\bar\theta_{23}}{Z_{23}}\right)
$$
$$
\DD^{-}_2 \equiv D^{-}_2 +
A\left(\frac{\theta_{21}}{Z_{21}}-
\frac{\theta_{23}}{Z_{23}}\right)+
(A+B)\frac{Z^{+}_{34}}{Z^{+}_{32}}
\left(\frac{\theta_{43}}{Z_{43}}-
\frac{\theta_{14}}{Z_{14}}\right)
$$
$$
\DD_3 \equiv\dd_3-
\frac{\bar\theta_{32}}{Z_{32}}D^{-}_3-
\frac{\theta_{34}}{Z_{34}}D^{+}_3 
\ ;\ \DD^{+}_3 \equiv D^{+}_3 -
B\left(\frac{\bar\theta_{34}}{Z_{34}}-
\frac{\bar\theta_{32}}{Z_{32}}\right)
$$
$$
\DD^{-}_3 \equiv D^{-}_3 +
B\left(\frac{\theta_{34}}{Z_{34}}-
\frac{\theta_{32}}{Z_{32}}\right)+
(A+B)\frac{Z^{+}_{21}}{Z^{+}_{23}}
\left(\frac{\theta_{12}}{Z_{12}}-
\frac{\theta_{14}}{Z_{14}}\right)
$$
$$
\DD_4 \equiv\dd_4-
\frac{\bar\theta_{43}}{Z_{43}}D^{-}_4-
\frac{\theta_{41}}{Z_{41}}D^{+}_4 
\ ;\ \DD^{-}_4 \equiv D^{-}_4 -
B\left(\frac{\theta_{43}}{Z_{43}}-
\frac{\theta_{41}}{Z_{41}}\right)
$$
$$
\DD^{+}_4 \equiv D^{+}_4 +
B\left(\frac{\bar\theta_{43}}{Z_{43}}-
\frac{\bar\theta_{41}}{Z_{41}}\right)+
(A+B)\frac{Z^{-}_{12}}{Z^{-}_{14}}
\left(\frac{\bar\theta_{12}}{Z_{12}}-
\frac{\bar\theta_{23}}{Z_{23}}\right)
$$ 
where
$$
A\equiv \frac{\u-\ell_1-\ell_2}{2} - \alpha
\ ;\ B\equiv \frac{\u+\ell_1+\ell_2}{2} +b_2-b_1+\alpha
\ ;\ A+B = \u+b_2-b_1
$$
The common solution of these twelve 
equations~$\DD_i\RR=0, \DD^{\pm}_i\RR=0\ i=1,2,3,4$ 
is the manifestly superconformal-invariant 
function~(see Appendix A):
\be
R(\vec Z) = 
\left(1 + A\frac{\theta^3_{12}\bar\theta^3_{12}}{Z^3_{12}}\right) 
\cdot\left(1 + B\frac{\theta^3_{14}\bar\theta^3_{14}}{Z^3_{14}}\right) 
- (A + B)\frac{\theta^3_{14}\bar\theta^3_{12}}{Z^3_{14}}=
\label{answ}
\ee
$$
=\left(1 + \frac{\theta^3_{12}\bar\theta^3_{12}}{Z^3_{12}}-
\frac{\theta^3_{14}\bar\theta^3_{12}}{Z^3_{14}}\right)^A 
\cdot\left(1 + \frac{\theta^3_{14}\bar\theta^3_{14}}{Z^3_{14}}-
\frac{\theta^3_{14}\bar\theta^3_{12}}{Z^3_{14}}\right)^B 
$$
where 
$$
Z^k_{ij}\equiv\frac{Z_{ij}}{Z_{ik}Z_{jk}}
\ ,\  \theta^k_{ij}\equiv
\frac{\theta_{ik}}{Z_{ik}}-\frac{\theta_{jk}}{Z_{jk}} 
\ ,\  \bar\theta^k_{ij}\equiv
\frac{\bar\theta_{ik}}{Z_{ik}}-\frac{\bar\theta_{jk}}{Z_{jk}}
$$
\end{prop}
This proposition can be proved by direct calculations.
It is remarkable that all involved operators $\DD_i\ ,\ \DD^{\pm}_i$
can be obtained by the similarity transformation from the 
simplest derivatives $\dd_i, D^{\pm}_i$
$$
\DD_i = R \dd_i R^{-1}\ ;\ \DD^{\pm}_i = R D^{\pm}_i R^{-1}
\ ;\ R = R(\vec Z).
$$

\subsection{The integral representation.}

It remains to fix the contours of integrations.
The $\R$-operator acts on the space of polynomials 
$\psi(Z_1,Z_2)$ which can be of arbitrary large degree.
The requirement of convergence allows the 
compact contours only.
The final representation for the R-matrix is: 
\be
\left[\R_{\vec\ell_1,\vec\ell_2}(u)\psi\right](Z_1,Z_2)
= \rho \int_{z_1}^{z_2}\rmd Z_3 
\int_{z_1}^{z_3}\rmd Z_4 \RR(Z_1,Z_2|Z_3,Z_4)\psi(Z_3,Z_4)
\label{SR1}
\ee  
where $Z\equiv\{z,\theta,\bar\theta\}$ 
$$
\int_{z_1}^{z_2}\rmd Z_3 \equiv 
\int_{z_1}^{z_2} \rmd z_3 \int \rmd \theta_3 \int \rmd \bar\theta_3. 
\ ;\  \int_{z_1}^{z_3}\rmd Z_4 \equiv 
\int_{z_1}^{z_3} \rmd z_4 \int \rmd \theta_4 \int \rmd \bar\theta_4
$$
and 
$$
\RR(\vec Z) = 
Z_{12}^{a}Z_{14}^{b}
Z_{32}^{c}Z_{34}^{d}\cdot
\left(1+\frac{\theta_{12}\bar\theta_{12}}{Z_{12}}\right)^{\alpha}
\left(1+\frac{\theta_{14}\bar\theta_{14}}{Z_{14}}\right)^{\beta}
\left(1+\frac{\theta_{32}\bar\theta_{32}}{Z_{32}}\right)^{\gamma}
\left(1+\frac{\theta_{34}\bar\theta_{34}}{Z_{34}}\right)^{\delta}
\cdot
$$
\be
\cdot
\left\{\left(1 + A\frac{\theta^3_{12}\bar\theta^3_{12}}{Z^3_{12}}\right) 
\cdot\left(1 + B\frac{\theta^3_{14}\bar\theta^3_{14}}{Z^3_{14}}\right) 
- (A + B)\frac{\theta^3_{14}\bar\theta^3_{12}}{Z^3_{14}}\right\}
\label{kernel}
\ee
$$
a\equiv \u-\ell_1-\ell_2
\ ;\  b\equiv -\u-\ell_1+\ell_2
\ ;\  c\equiv -\u+\ell_1-\ell_2
\ ;\  d\equiv \u+\ell_1+\ell_2 
$$
$$
\beta=b_1-\alpha
\ ;\ \gamma = -b_2-\alpha
\ ;\ \delta = b_2-b_1+\alpha. 
$$
The normalisation factor $\rho$ 
is chosen so that $\R: 1\mapsto 1$
\be
\rho =
\frac{\Gamma(c+b+d)}{\Gamma(b)\Gamma(c)\Gamma(d)} 
\frac{(\u-\ell_1+b_2)(\u-\ell_2-b_1)(\u+\ell_1+\ell_2)}
{-\u+\ell_1+\ell_2}
\label{snorm} 
\ee
It is possible to check $a\ posteriori$ that 
the obtained integral operator 
has the matrix elements~(\ref{ABCD}).
Using the formulae from Appendix B we derive the 
formulae for the even basis vectors:
$$
\int_{z_1}^{z_2}\rmd Z_3 
\int_{z_1}^{z_3}\rmd Z_4 \RR(Z_1,Z_2|Z_3,Z_4)
\left(Z_{34}+\frac{\theta_{34}\bar\theta_{34}}{2}\right)^n=
$$
$$
=\frac{\Gamma(b)\Gamma(c)\Gamma(d+n)}{\Gamma(c+b+d+n)}\cdot
\left\{\RR_{++}
\left(Z_{21}+\frac{\theta_{21}\bar\theta_{21}}{2}\right)^n+
\RR_{-+}
\left(Z_{21}-\frac{\theta_{21}\bar\theta_{21}}{2}\right)^n
\right\}
$$
where 
$$
\RR_{++}= (\ell_1+b_1)(\ell_2-b_2)
\ ;\ \RR_{-+}= \frac{\u^3+\u^2(b_2-b_1)-\u(\ell_1^2+\ell_2^2+2b_1b_2)+
(b_1+b_2)(\ell_1^2-\ell_2^2)}{-\u+\ell_1+\ell_2+n}
$$
and
$$
\int_{z_1}^{z_2}\rmd Z_3 
\int_{z_1}^{z_3}\rmd Z_4 \RR(Z_1,Z_2|Z_3,Z_4)
\left(Z_{34}-\frac{\theta_{34}\bar\theta_{34}}{2}\right)^n=
$$
$$
=\frac{\Gamma(b)\Gamma(c)\Gamma(d+n)}{\Gamma(c+b+d+n)}\cdot
\left\{\RR_{+-}
\left(Z_{21}+\frac{\theta_{21}\bar\theta_{21}}{2}\right)^n+
\RR_{--}
\left(Z_{21}-\frac{\theta_{21}\bar\theta_{21}}{2}\right)^n
\right\}
$$
where
$$
\RR_{+-}= -(\u+b_2-b_1)(\u+\ell_1+\ell_2+n)
\ ;\ \RR_{--} = 
\frac{(\ell_1-b_1)(\ell_2+b_2)(\u+\ell_1+\ell_2+n)}
{-\u+\ell_1+\ell_2+n} 
$$
$$
\left.\RR_{++}+\RR_{-+}\right|_{n=0}=
\left.\RR_{+-}+\RR_{--}\right|_{n=0}=
\frac{-\u+\ell_1+\ell_2}
{(\u-\ell_1+b_2)(\u-\ell_2-b_1)(\u+\ell_1+\ell_2)}
$$
The analogous formulae for the odd vectors are 
much simpler: 
$$
\int_{z_1}^{z_2}\rmd Z_3 
\int_{z_1}^{z_3}\rmd Z_4 \RR(Z_1,Z_2|Z_3,Z_4)
Z^{n}_{34}\cdot
\left(\begin{array}{c}
\theta_{34} \\ \bar\theta_{34}   
\end{array} \right )=
$$
$$
=-\frac{\Gamma(b)\Gamma(c)\Gamma(d+n+1)}{\Gamma(c+b+d+n+1)}
Z_{21}^{n}\cdot
\left(\begin{array}{c}
(\u+b_2+\ell_1)(\u+b_2-\ell_1)\theta_{21} \\ 
(\u-b_1-\ell_2)(\u-b_1+\ell_2)\bar\theta_{21}   
\end{array} \right )
$$
These formulae for the action of the R-matrix on the 
basis vectors coincide up to overall normalisation 
with the ones~(\ref{ABCD}) obtained in our previous paper~\cite{DKK}.

\section{Conclusions}

We have derived the integral operator form of the rational
$sl(2|1)$ symmetric solution of the Yang-Baxter equation 
acting on the tensor product of two general 
lowest weight representations
$\vec\ell_1 =(\ell_1, b_1)$ and 
$\vec\ell_2 =(\ell_2, b_2)$. The lowest weight
representations space is identified with the space of polynomials in one
even $(z)$ and two odd $(\theta, \bar\theta )$. The kernel of the 
R-operator is a superconformal four-point function of primary fields with
weights $\vec\ell_1 ,\vec\ell_2 $
and $\vec\ell_3 = -\vec\ell_1
\ ,\  \vec\ell_4 = - \vec\ell_2 $.
This function of the supercoordinates of  four points arises as a simple
sum of terms in powers of $Z_{ij}, \theta_{ij}, \bar\theta_{ij} $.
The simplicity of the result makes this integral operator form useful in
applications and further investigations.

Following the well known procedure, the general R-operator is derived from
the Yang-Baxter equation involving the latter and the known R-operator 
$R_{f, \ell}$ acting in the tensor product of the fundamental representation
(in matrix form) and an arbitrary representation. We have formulated the
conditions involved in this Yang-Baxter relation as differential equations
on the kernel.

There is a remarkable reductions of these equations, 
being an overdetermined
system of second order equations, to a simple set of first order equations. 
In this way the obtained kernel expresses the similarity 
transformation relating this consistent system of first 
order equations to the trivial one being the condition 
for a function to be constant in all variables~$z_i,\theta_i,\bar\theta_i$. 
We have obtained this reduction by observing that the Lax matrix
representing $R_{f, \ell}$ has a peculiar structure implying that
appropriate projections reduce the number of derivatives.

We have calculated the matrix elements of the obtained R-operator.
The result is consistent with the one obtained earlier from
recurrence relations in the framework of the matrix formulation.
We have quoted the  formulae to be used for calculating the 
action of the integral operator.

The method of calculation relying on this reduction seems to be general
enough to become useful for solving the Yang-Baxter equations with other
symmetries. The structure on which the reduction relies deserves
further study.

\section{Acknowledgments}

We thank G.Korchemsky and A.Manashov for the stimulating 
discussion and critical remarks.
This work has been supported by 
Deutsche Forschungsgemeinschaft, grant No  KI 623/1-2
and by INTAS,grant No 96-524. 
One of us (D.K.) is grateful to 
the German Ministry BMBF for support.
The work of S.D. was partially supported by 
RFFI-grant No 00-01-005000. 

\section*{Appendix A} 
 
In this Appendix we shall construct the superconformal invariant 
N-point function with zero global U(1)-charge.
The requirement of $s\ell(2|1)$-invariance~(\ref{sinvar1}) 
results in the equations:
$$
V^{-}\RR(\vec Z) = 0\ ;\ W^{-}\RR(\vec Z) = 0\ ;\ 
V^{+}\RR(\vec Z) = 0\ ;\ W^{+}\RR(\vec Z) = 0  
$$
where $V^{-}=\sum_{i=1}^{4} V^{-}_i$ and so on. 
Note we show the independent equations only.
The others follow from the commutation relations  
$$
\{V^{\pm},W^{\pm}\} =\pm S^{\pm}
\ ;\  \{V^{\pm},W^{\mp}\} = - S \pm B,
$$
Let us consider the general case of N-point functions 
with zero total $U(1)$-charge:
$\vec Z = (Z_1,...,Z_N)\ ;\ \sum_{i} b_i = 0$.
It is convenient to use the global 
$s\ell(2|1)$-transformations which are generated 
by the lowering operators~$V^{-},W^{-}$
$$
e^{\epsilon V^{-}}\Phi(z;\theta,\bar\theta)=
\Phi\left(z+\frac{\epsilon\bar\theta}{2};\theta+\epsilon,\bar\theta\right)
\ ;\ e^{\epsilon W^{-}}\Phi(z;\theta,\bar\theta)=
\Phi\left(z+\frac{\epsilon\theta}{2};\theta,\bar\theta+\epsilon\right).
$$
and the rising operators~$V^{+},W^{+}$ 
$$
e^{\epsilon V^{+}}\Phi(z;\theta,\bar\theta)=
\frac{1}{(1+\epsilon\bar\theta)^{\ell-b}}
\Phi\left(\frac{z}{1+\frac{\epsilon \bar\theta}{2}};
\frac{\theta -\epsilon z}
{1+\frac{\epsilon \bar\theta}{2}} ,\bar\theta\right) ,
$$ 
$$
e^{\epsilon W^{+}}\Phi(z;\theta,\bar\theta)=
\frac{1}{(1+\epsilon \theta)^{\ell+b}}
\Phi\left(\frac{z}{1+\frac{\epsilon \theta}{2}};\theta,
\frac{\bar\theta -\epsilon z}{1+\frac{\epsilon \theta}{2}}\right),
$$ 
so that the requirement of superconformal 
invariance results in the system of equations:
$$
V^{-}:\ \ \RR(Z_1...Z_N) = 
\RR\left(z_1+\frac{\epsilon\bar\theta_1}{2},
\theta_1+\epsilon,\bar\theta_1...
z_N+\frac{\epsilon\bar\theta_N}{2},
\theta_N+\epsilon,\bar\theta_N \right)
$$
$$
W^{-}:\ \ \RR(Z_1...Z_N) = 
\RR\left(z_1+\frac{\epsilon\theta_1}{2},\theta_1,\bar\theta_1+\epsilon
... 
z_N+\frac{\epsilon\theta_N}{2},\theta_N,\bar\theta_N+\epsilon
\right)
$$
$$
V^{+}:\ \ \RR(Z_1...Z_N)=
\prod_{i=1}^N 
\frac{1}{(1+\epsilon\bar\theta_i)^{\ell_i-b_i}}
\RR\left(\frac{z_1}{1+\frac{\epsilon \bar\theta_1}{2}},
\frac{\theta_1 -\epsilon z_1}
{1+\frac{\epsilon \bar\theta_1}{2}} ,\bar\theta_1 ...
\frac{z_N}{1+\frac{\epsilon \bar\theta_N}{2}},
\frac{\theta_N -\epsilon z_N}
{1+\frac{\epsilon \bar\theta_N}{2}} ,\bar\theta_N\right) ,
$$ 
$$
W^{+}:\ \ \RR(Z_1...Z_N)=
\prod_{i=1}^N 
\frac{1}{(1+\epsilon \theta_i)^{\ell_i+b_i}}
\RR\left(\frac{z_1}{1+\frac{\epsilon \theta_1}{2}},\theta_1,
\frac{\bar\theta_1 -\epsilon z_1}{1+\frac{\epsilon \theta_1}{2}} ...
\frac{z_N}{1+\frac{\epsilon \theta_N}{2}},\theta_N,
\frac{\bar\theta_N -\epsilon z_N}{1+\frac{\epsilon \theta_N}{2}}
\right).
$$
Solutions of $V^{-}$- and $W^{-}$-equations are
shift-invariant functions which depend 
on the superdifferences only
$$
\RR(\vec Z) = R(Z_{ij}, \theta_{ij},\bar\theta_{ij}),
$$   
$$
Z_{ij}\equiv 
z_i-z_j+\frac{1}{2}(\bar\theta_i\theta_j+\theta_i\bar\theta_j)
\ ;\ \theta_{ij}\equiv \theta_i - \theta_j
\ ;\ \bar\theta_{ij}\equiv \bar\theta_i - \bar\theta_j.
$$
Let us fix some point $k$. 
There are $N-1$ independent even 
differences $Z_{ik}, i\neq k$ for fixed $k$ and 
$2(N-1)$ independent odd differences 
$\theta_{ik},\bar\theta_{ik}\ ,\ i\neq k$ that 
transform with respect of the remaining 
transformations as follows 
$$
V^{+}:\ \  
Z_{ij}\to \frac{Z_{ij}}{(1+\frac{\epsilon \bar\theta_i}{2})
(1+\frac{\epsilon \bar\theta_j}{2})}
\ ;\ \theta_{ij}\to \frac{\theta_{ij}-\epsilon Z_{ij}}
{(1+\frac{\epsilon \bar\theta_i}{2})
(1+\frac{\epsilon \bar\theta_j}{2})}
\ ;\ \bar\theta_{ij}\to \bar\theta_{ij}
$$
$$
W^{+}:\ \  
Z_{ij}\to \frac{Z_{ij}}{(1+\frac{\epsilon \theta_i}{2})
(1+\frac{\epsilon \theta_j}{2})}
\ ;\ \bar\theta_{ij}\to \frac{\bar\theta_{ij}-\epsilon Z_{ij}}
{(1+\frac{\epsilon \theta_i}{2})
(1+\frac{\epsilon \theta_j}{2})}
\ ;\ \theta_{ij}\to \theta_{ij}.
$$
It is useful to extract the 
representation dependent part from the 
function $\RR(\vec Z)$
$$
\RR(\vec Z) = \prod_{i<j} Z_{ij}^{a_{ij}}
\left(1+\frac{\theta_{ij}\bar\theta_{ij}}{Z_{ij}}\right)^{b_{ij}}
R(\vec Z).
$$
Using the formulae 
$$
V^{+}:\ \ 
\frac{\theta_{ij}}{Z_{ij}}\to \frac{\theta_{ij}}{Z_{ij}}-\epsilon 
\ ;\ 1+\frac{\theta_{ij}\bar\theta_{ij}}{Z_{ij}}\to 
\left(1+\frac{\theta_{ij}\bar\theta_{ij}}{Z_{ij}}\right)
(1-\epsilon\bar\theta_{ij})
$$
$$
W^{+}:\ \  
\frac{\bar\theta_{ij}}{Z_{ij}}\to \frac{\bar\theta_{ij}}{Z_{ij}}-\epsilon 
\ ;\ 1+\frac{\theta_{ij}\bar\theta_{ij}}{Z_{ij}}\to 
\left(1+\frac{\theta_{ij}\bar\theta_{ij}}{Z_{ij}}\right)
(1+\epsilon\theta_{ij})
$$
it is easily to check that under the conditions
$$
\sum_{i\neq j}a_{ij} = - 2\ell_j\ ,\ a_{ij} = a_{ji}
\ ;\ \sum_{i\neq j}b_{ij} = - b_j
\ ,\ b_{ij} = -b_{ji}
$$
the equations for the function $R(\vec Z)$ reads as follows 
$$
V^{+}:\ \ R(Z_1...Z_N)=
R\left(\frac{z_1}{1+\frac{\epsilon \bar\theta_1}{2}},
\frac{\theta_1 -\epsilon z_1}
{1+\frac{\epsilon \bar\theta_1}{2}} ,\bar\theta_1 ...
\frac{z_N}{1+\frac{\epsilon \bar\theta_N}{2}},
\frac{\theta_N -\epsilon z_N}
{1+\frac{\epsilon \bar\theta_N}{2}} ,\bar\theta_N\right) ,
$$ 
$$
W^{+}:\ \ R(Z_1...Z_N)=
R\left(\frac{z_1}{1+\frac{\epsilon \theta_1}{2}},\theta_1,
\frac{\bar\theta_1 -\epsilon z_1}{1+\frac{\epsilon \theta_1}{2}} ...
\frac{z_N}{1+\frac{\epsilon \theta_N}{2}},\theta_N,
\frac{\bar\theta_N -\epsilon z_N}{1+\frac{\epsilon \theta_N}{2}}
\right),
$$   
i.e. the function $R(\vec Z)$ is the general function 
of all possible superconformal invariants.
Let us go to the construction of the 
superconformal invariants.
For this purpose we fix the two points $Z_j, Z_k$ and 
choose the new variables~($j$ and $k$ are fixed) 
$$
Z^k_{ij}\equiv\frac{Z_{ji}}{Z_{ik}Z_{jk}}
\ ,\  \theta^k_{ij}\equiv
\frac{\theta_{ik}}{Z_{ik}}-\frac{\theta_{jk}}{Z_{jk}} 
\ ,\  \bar\theta^k_{ij}\equiv
\frac{\bar\theta_{ik}}{Z_{ik}}-\frac{\bar\theta_{jk}}{Z_{jk}}
$$
with the simple transformation properties 
$$
V^{+}:\ \  
Z^k_{ij}\rightarrow Z^k_{ij}\cdot(1+\epsilon \bar\theta_k)
\ ;\ \theta^k_{ij}\rightarrow \theta^k_{ij}
\ ;\ \bar\theta^k_{ij}\rightarrow 
\bar\theta^k_{ij}\cdot(1+\epsilon \bar\theta_k)
$$
$$
W^{+}:\ \  
Z^k_{ij}\rightarrow Z^k_{ij}\cdot(1+\epsilon \theta_k)
\ ;\ \theta^k_{ij}\rightarrow 
\theta^k_{ij}\cdot(1+\epsilon \theta_k)
\ ;\ \bar\theta^k_{ij}\rightarrow \bar\theta^k_{ij}.
$$
It is easy to construct the superconformal 
invariants using these variables.
Let us fix the three points~$Z_j, Z_k, Z_p$. 
Then the set of invariants is the following: 
$$
\frac{Z^k_{ij}}{Z^k_{pj}} = \frac{Z_{ij}Z_{pk}}{Z_{ik}Z_{pj}}
\ ;\ \frac{\theta^k_{ij}\bar\theta^k_{nm}}{Z^k_{pj}}
$$
Finally we obtain the general solution of conformal constraints
\be
\RR(\vec z) = 
\prod_{i<j} z_{ij}^{a_{ij}}
\left(1+\frac{\theta_{ij}\bar\theta_{ij}}{Z_{ij}}\right)^{b_{ij}}  
R\left(\frac{Z_{ij}Z_{pk}}{Z_{ik}Z_{pj}};
\frac{\theta^k_{ij}\bar\theta^k_{nm}}{Z^k_{pj}}
\right)
\label{CC}
\ee 
$$
\sum_{i\neq j}a_{ij} = - 2\ell_j\ ,\ a_{ij} = a_{ji}
\ ;\ \sum_{i\neq j}b_{ij} = - b_j
\ ,\ b_{ij} = -b_{ji}
$$
where $j,k,p$ are fixed.
In the case of four-point function we can choose the 
superconformal invariant ansatz in the form
$$
\RR(\vec Z) = 
Z_{12}^{a}Z_{14}^{b}
Z_{32}^{c}Z_{34}^{d}\cdot
\left(1+\frac{\theta_{12}\bar\theta_{12}}{Z_{12}}\right)^{\alpha}
\left(1+\frac{\theta_{14}\bar\theta_{14}}{Z_{14}}\right)^{\beta}
\left(1+\frac{\theta_{32}\bar\theta_{32}}{Z_{32}}\right)^{\gamma}
\left(1+\frac{\theta_{34}\bar\theta_{34}}{Z_{34}}\right)^{\delta}\cdot
$$
$$
\cdot R\left(\frac{Z_{12}Z_{34}}{Z_{13}Z_{24}};
\frac{\theta^3_{12}\bar\theta^3_{12}}{Z^3_{12}} ,
\frac{\theta^3_{14}\bar\theta^3_{14}}{Z^3_{14}} ,
\frac{\theta^3_{12}\bar\theta^3_{14}}{Z^3_{14}} ,
\frac{\theta^3_{14}\bar\theta^3_{12}}{Z^3_{12}}\right)
$$
where
$$
a\equiv \u-\ell_1-\ell_2
\ ;\  b\equiv -\u-\ell_1+\ell_2
\ ;\  c\equiv -\u+\ell_1-\ell_2
\ ;\  d\equiv \u+\ell_1+\ell_2 
$$
$$
\beta=b_1-\alpha
\ ;\ \gamma = -b_2-\alpha
\ ;\ \delta = b_2-b_1+\alpha. 
$$

\section*{Appendix B} 

In this Appendix we collect all needed integration formulae.\\
\underline{{\bf General formulae of integration}}
$$
\int_{z_4}^{z_2}\rmd Z_3 
Z_{23}^{A}Z_{34}^{B} 
\left(1+\frac{\theta_{23}\bar\theta_{23}}{Z_{23}}\right)^a
\left(1+\frac{\theta_{34}\bar\theta_{34}}{Z_{34}}\right)^b=
-\frac{\Gamma(A)\Gamma(B)}{\Gamma(A+B)}
\left[\frac{A b+B a}{A+B}+
\left(ab+\frac{AB}{4}\right)
\frac{\theta_{24}\bar\theta_{24}}{Z_{24}}\right]
Z_{24}^{A+B}
$$
$$
\int_{z_4}^{z_2}\rmd Z_3 
Z_{23}^{A}Z_{34}^{B}
\cdot
\left(\begin{array}{c}
\theta_{34} \\ \bar\theta_{34}   
\end{array} \right ) =
\half\frac{\Gamma(A+1)\Gamma(B+1)}{\Gamma(A+B+1)}
Z_{24}^{A+B}\cdot
\left(\begin{array}{c}
\theta_{24} \\ -\bar\theta_{24}   
\end{array} \right )
$$
$$
\int_{z_4}^{z_2}\rmd Z_3 
Z_{23}^{A}Z_{34}^{B}
\cdot
\left(\begin{array}{c}
\theta_{23} \\ \bar\theta_{23}   
\end{array} \right )=
-\half\frac{\Gamma(A+1)\Gamma(B+1)}{\Gamma(A+B+1)}
Z_{24}^{A+B}\cdot
\left(\begin{array}{c}
\theta_{24} \\ -\bar\theta_{24}   
\end{array} \right )
$$
\underline{{\bf Formulae of integration for even vectors}}
\begin{itemize}
\item
$$
\int_{z_1}^{z_3}\rmd Z_4 
Z_{34}^{d}Z_{41}^{b} 
\left(1+\frac{\theta_{14}\bar\theta_{14}}{Z_{14}}\right)^{\beta}
\left(1+\frac{\theta_{34}\bar\theta_{34}}{Z_{34}}\right)^{\delta}
\left\{1+B\frac{\theta^3_{14}\bar\theta^3_{14}}{Z^3_{14}}\right\} 
Z^{n}_{34}\left(1+J\frac{\theta_{34}\bar\theta_{34}}{Z_{34}}\right)
=
$$
$$
=\frac{\Gamma(b)\Gamma(d+n)}{\Gamma(b+d+n)}Z_{31}^{b+d+n}
\left\{C_1 + C_2\frac{\theta_{31}\bar\theta_{31}}{Z_{31}}\right\}
$$
$$
C_1=B+\frac{\beta(d+n)-b(\delta+J)}{b+d+n}
\ ;\ C_2 = \beta(\delta+J)-\frac{b(d+n)}{4}+
B\left(\delta-\beta+J+\frac{\beta(d+n)-b(\delta+J)}{b+d+n}\right) 
$$
\item
$$
Z^a_{21}\left(1+\frac{\theta_{12}\bar\theta_{12}}{Z_{12}}\right)^{\alpha}
\int_{z_1}^{z_2}\rmd Z_3 
Z_{23}^{c} 
\left(1+\frac{\theta_{32}\bar\theta_{32}}{Z_{32}}\right)^{\gamma}
\left\{1+A\frac{\theta^3_{14}\bar\theta^3_{14}}{Z^3_{14}}\right\} 
Z^{b+d+n}_{31}
\left(1+J\frac{\theta_{31}\bar\theta_{31}}{Z_{31}}\right)
=
$$
$$
=-\frac{\Gamma(c)\Gamma(b+d+n)}{\Gamma(c+b+d+n)}
Z_{21}^{a+b+c+d+n}
\left\{C_1 + C_2\frac{\theta_{21}\bar\theta_{21}}{Z_{21}}\right\}
$$
$$
C_1 = A+\frac{c J-\gamma(b+d+n)}{b+c+d+n} ,
$$
$$
C_2 = -\gamma J+\frac{c(b+d+n)}{4}-
\alpha\frac{c J-\gamma(b+d+n)}{b+c+d+n}+
A\left(J-\gamma-\alpha-\frac{c J-\gamma(b+d+n)}{b+c+d+n}\right) 
$$
\item
$$
Z^a_{21}\left(1+\frac{\theta_{12}\bar\theta_{12}}{Z_{12}}\right)^{\alpha}
\int_{z_1}^{z_2}\rmd Z_3 
Z_{23}^{c} 
\left(1+\frac{\theta_{32}\bar\theta_{32}}{Z_{32}}\right)^{\gamma}\cdot
$$
$$
\cdot\int_{z_1}^{z_3}\rmd Z_4 
Z_{34}^{d}Z_{41}^{b} 
\left(1+\frac{\theta_{14}\bar\theta_{14}}{Z_{14}}\right)^{\beta}
\left(1+\frac{\theta_{34}\bar\theta_{34}}{Z_{34}}\right)^{\delta}
\frac{\theta^3_{14}\bar\theta^3_{12}}{Z^3_{14}} 
Z^{n}_{34}\left(1+J\frac{\theta_{34}\bar\theta_{34}}{Z_{34}}\right)=
$$
$$
=-\frac{\Gamma(b)\Gamma(d+n)}{\Gamma(b+d+n)}
\frac{\Gamma(c)\Gamma(b+d+n)}{\Gamma(c+b+d+n)}
Z_{21}^{a+b+c+d+n} C_1\cdot
\left\{1 + C_2 \frac{\theta_{21}\bar\theta_{21}}{Z_{21}} \right\}
$$
$$
C_1 = \delta+\beta+J-\frac{b+d+n}{2}
\ ;\ C_2 = \frac{b+d+n}{2}-\alpha-\gamma 
$$
\end{itemize}
\underline{{\bf Formulae of integration for odd vectors}}
\begin{itemize}
\item
$$
\int_{z_1}^{z_3}\rmd Z_4 
Z_{34}^{d}Z_{41}^{b} 
\left(1+\frac{\theta_{14}\bar\theta_{14}}{Z_{14}}\right)^{\beta}
\left(1+\frac{\theta_{34}\bar\theta_{34}}{Z_{34}}\right)^{\delta}
\left\{1+B\frac{\theta^3_{14}\bar\theta^3_{14}}{Z^3_{14}}\right\} 
Z^{n}_{34}\cdot
\left(\begin{array}{c}
\theta_{34} \\ \bar\theta_{34}   
\end{array} \right )
=
$$
$$
=\frac{\Gamma(b)\Gamma(d+n+1)}{\Gamma(b+d+n+1)}Z_{31}^{b+d+n}
\cdot
\left(\begin{array}{c}
\left(\beta-\frac{b}{2}+B\right)\theta_{31} \\ 
\left(\beta+\frac{b}{2}+B\right)\bar\theta_{31}   
\end{array} \right )
$$
\item
$$
Z^a_{21}\left(1+\frac{\theta_{12}\bar\theta_{12}}{Z_{12}}\right)^{\alpha}
\int_{z_1}^{z_2}\rmd Z_3 
Z_{23}^{c} 
\left(1+\frac{\theta_{32}\bar\theta_{32}}{Z_{32}}\right)^{\gamma}
\left\{1+A\frac{\theta^3_{14}\bar\theta^3_{14}}{Z^3_{14}}\right\} 
Z^{b+d+n}_{31}\cdot
\left(\begin{array}{c}
\theta_{31} \\ \bar\theta_{31}   
\end{array} \right )
=
$$
$$
=\frac{\Gamma(c)\Gamma(b+d+n+1)}{\Gamma(c+b+d+n+1)}
Z_{21}^{a+b+c+d+n}\cdot
\left(\begin{array}{c}
\left(\gamma+\frac{c}{2}-A\right)\theta_{21} \\ 
\left(\gamma-\frac{c}{2}-A\right)\bar\theta_{21}   
\end{array} \right )
$$
\item
$$
Z^a_{21}\left(1+\frac{\theta_{12}\bar\theta_{12}}{Z_{12}}\right)^{\alpha}
\int_{z_1}^{z_2}\rmd Z_3 
Z_{23}^{c} 
\left(1+\frac{\theta_{32}\bar\theta_{32}}{Z_{32}}\right)^{\gamma}
\cdot
$$
$$
\cdot\int_{z_1}^{z_3}\rmd Z_4 
Z_{34}^{d}Z_{41}^{b} 
\left(1+\frac{\theta_{14}\bar\theta_{14}}{Z_{14}}\right)^{\beta}
\left(1+\frac{\theta_{34}\bar\theta_{34}}{Z_{34}}\right)^{\delta}
\frac{\theta^3_{14}\bar\theta^3_{12}}{Z^3_{14}} 
Z^{n}_{34}\cdot
\left(\begin{array}{c}
\theta_{34} \\ \bar\theta_{34}   
\end{array} \right )=
$$
$$
=\frac{\Gamma(b)\Gamma(d+n+1)}{\Gamma(b+d+n+1)}
\frac{\Gamma(c)\Gamma(b+d+n+1)}{\Gamma(c+b+d+n+1)}
Z_{21}^{a+b+c+d+n}\cdot
\left(\begin{array}{c}
0 \\ \left(\gamma-\beta-\frac{b+c}{2}\right)\bar\theta_{21}   
\end{array} \right )
$$
\end{itemize}

\end{document}